\documentclass[twocolumn]{aastex701}

\usepackage{amsmath}

\begin{document}
\title{B-fields And dust in interstelLar fiLAments using Dust POLarization (BALLAD-POL): IV. Grain alignment mechanisms in Cocoon Nebula (IC 5146) using polarization observations from JCMT/POL-2}

\author[0009-0002-6171-9740]{Saikhom Pravash}
\affiliation{Indian Institute of Astrophysics, II Block, Koramangala, 560034, India}
\affiliation{Pondicherry University, R.V. Nagar, Kalapet, Puducherry, 605014, India}
\email[show]{\href{mailto:saikhom.singh@iiap.res.in, spravash11@gmail.com}{saikhom.singh@iiap.res.in, spravash11@gmail.com}}
%\email[show]{\href{mailto:spravash11@gmail.com{spravash11@gmail.com}}
%\affiliation{Pondicherry University, R.V. Nagar, Kalapet, Puducherry, 605014, India}

\author[0000-0003-2017-0982]{Thiem Hoang}
\affiliation{Korea Astronomy and Space Science Institute, 776 Daedeokdae-ro, Yuseong-gu, Daejeon 34055, Republic of Korea}
\affiliation{University of Science and Technology, Korea, 217 Gajeong-ro, Yuseong-gu, Daejeon 34113, Republic of Korea}
\email{thiemhoang@kasi.re.kr}

\author[0000-0002-6386-2906]{Archana Soam}
\affiliation{Indian Institute of Astrophysics, II Block, Koramangala, 560034, India}
\email{archana.soam@iiap.res.in}

\author[0000-0003-0014-1527]{Eun Jung Chung}
\affiliation{Korea Astronomy and Space Science Institute, 776 Daedeokdae-ro, Yuseong-gu, Daejeon 34055, Republic of Korea}
\affiliation{Department of Astronomy and Space Science, Chungnam National University, Daejeon, Republic of Korea}
%\affiliation{Korea Astronomy and Space Science Institute, 776 Daedeokdae-ro, Yuseong-gu, Daejeon 34055, Republic of Korea}
\email{rigelej@gmail.com}

\author[0000-0002-2808-0888]{Pham Ngoc Diep}
\affiliation{Department of Astrophysics, Vietnam National Space Center, Vietnam Academy of Science and Technology, 18 Hoang Quoc Viet, Hanoi, Vietnam}
\affiliation{Graduate University of Science and Technology, Vietnam Academy of Science and Technology, 18 Hoang Quoc Viet, Hanoi, Vietnam}
\email{pndiep@vnsc.org.vn}

\author[0000-0002-5913-5554]{Nguyen Bich Ngoc}
\affiliation{Department of Astrophysics, Vietnam National Space Center, Vietnam Academy of Science and Technology, 18 Hoang Quoc Viet, Hanoi, Vietnam}
\affiliation{Graduate University of Science and Technology, Vietnam Academy of Science and Technology, 18 Hoang Quoc Viet, Hanoi, Vietnam}
\email{capi37capi@gmail.com}

\author[0000-0002-6488-8227]{Le Ngoc Tram}
\affiliation{Leiden Observatory, Leiden University, PO Box 9513, 2300 RA Leiden, The Netherlands}
\email{nle@mpifr-bonn.mpg.de}

\begin{abstract}
The polarization of starlight and thermal dust emission from aligned non-spherical grains provides a powerful tool for tracing magnetic field morphologies and strengths in diffuse interstellar medium to star-forming regions, and constraining dust grain properties and their alignment mechanisms. However, the physics of grain alignment is not yet fully understood. The alignment based on RAdiative Torques (RATs), known as RAT Alignment or RAT-A mechanism is the most acceptable mechanism. In this work, we investigate the grain alignment mechanisms in F13 (F13N and F13C) and F13S filamentary regions of the Cocoon Nebula (IC 5146) using polarized thermal dust emission observations from JCMT/POL-2 at 850 $\mu$m. We find that the polarization fraction decreases with increasing total intensity and gas column density in each region, termed as polarization hole. We investigate for any role of magnetic field tangling on the observed polarization hole by estimating the polarization angle dispersion function. Our study finds that the polarization hole is not significantly influenced by magnetic field tangling, but majorly due to decrease in RAT alignment efficiency of grains in denser regions. To test whether RAT-A mechanism can reproduce the observational results, we estimate minimum alignment size of grains using RAT theory. Our study finds strong evidence for RAT-A mechanism that can explain the polarization hole. We also find potential hints that the observed higher polarization fractions in some regions of F13 filament can be due to combined effects of both suprathermal rotation by RATs and enhanced magnetic relaxation, supporting the Magnetically-Enhanced RAT (M-RAT) mechanism.                
\end{abstract}


\section{Introduction} \label{section:Introduction}
Dust grains in the interstellar medium, molecular clouds, and the star-forming regions are very crucial elements that significantly influence \textbf{in} various \textbf{astrophysical} processes like star and planet formation; gas heating and cooling, and they provide the surface to form gas molecules (see \citealt{2003ARA&A..41..241D}). \cite{1949Sci...109..166H} and \cite{1949ApJ...109..471H} first observed the polarization of background starlight due to differential or selective or dichroic extinction by non-spherical grains aligned with the ambient interstellar magnetic fields. Since then, several studies try to explain the possible mechanisms for the alignment of the grains (e.g, \citealt{1951ApJ...114..206D}; \citealt{1979ApJ...231..404P}) (for a review see \citealt{2015ARA&A..53..501A}). However, these suggested mechanisms could not fully explain many observational results. The most acceptable mechanism of grain alignment that can explain various observational results in different environments from the diffuse interstellar medium, molecular clouds to the star-forming regions at different wavelengths from optical to sub-millimeter/millimeter is the alignment of grains based on RAdiative Torques (RATs), known as RAdiative Torque Alignment (RAT-A) mechanism. The RAT-A theory was first introduced by \cite{1976Ap&SS..43..291D} and numerically demonstrated in \cite{1997ApJ...480..633D}. An analytical RAT theory was later developed by \cite{2007MNRAS.378..910L} and \cite{2008MNRAS.388..117H}. According to RAT-A theory, grains of non-spherical in shape when exposed to external anisotropic radiation field experienced radiative torques which tend to increase the spin of the grains far above the thermal rotation rate to suprathermal rotation and align the grains with the ambient magnetic field (\citealt{1997ApJ...480..633D}; \citealt{2007MNRAS.378..910L}). The grains achieve efficient alignment when they rotate suprathermally at the rotation rate greater than about 3 times of their thermal angular velocity (\citealt{2008MNRAS.388..117H, 2016ApJ...831..159H}). 

The aligned dust grains also re-emit polarized thermal emission at longer wavelengths \citep{1988QJRAS..29..327H}. The dust polarization has been used widely to study the magnetic fields in various environments ranging from the diffuse interstellar medium to the star-forming regions in the molecular clouds and also to study the properties of the dust grains like shape, size, composition, etc. (e.g, \citealt{2021ApJ...919...65D}). Magnetic fields are thought to have significant influences on the formation and evolution of molecular clouds and regulation of processes of star formation (see \citealt{2012ARA&A..50...29C}; \citealt{2019FrASS...6...15P}). The alignment of the grains are in such a way that their short axes are parallel and the long axes are perpendicular to the ambient magnetic field (e.g, \citealt{2007JQSRT.106..225L}; \citealt{2015ARA&A..53..501A}; \citealt{2015psps.book...81L}). In the polarization of starlight due to dichroic extinction by aligned grains, the observed polarization angle traces the plane-of-sky (POS) projected magnetic fields. However, in the polarization of thermal dust emission, the observed polarization angle should be rotated by 90$^\circ$ to trace the POS projected magnetic fields (e.g, \citealt{1988QJRAS..29..327H}). In recent studies by \cite{2024ApJ...965..183H} and \cite{2024arXiv240714896T}, they suggest the potential of tracing three dimensional magnetic fields with full dust polarization data by comparing the observed dust polarization with the accurate model of dust polarization predicted from the grain alignment theory and dust properties. For the study of magnetic fields in the diffuse interstellar medium, molecular clouds and the star-forming regions with more precise using dust polarization, it is important to study the exact physical mechanisms for the alignment of grains.

According to the RAT-A mechanism, it is expected to have an anti-correlation of the polarization fraction with the total emission intensity or the gas column density and a correlation with the radiation field intensity or equivalently the dust temperatures (see \citealt{2020ApJ...896...44L}; \citealt{2021ApJ...908..218H}). A decreasing trend of the polarization fraction with the increase in the total emission intensity and the gas column density, usually termed as polarization hole, is observed in various studies at different regions (e.g, \citealt{2019FrASS...6...15P}; \citealt{2021ApJ...908..218H}). The polarization hole may be caused due to decrease in grain alignment efficiency in denser regions as expected by RAT-A theory or due to magnetic field fluctuations along the line-of-sight or both. The main cause of the magnetic field fluctuations is the turbulence in the molecular clouds (\citealt{1989ApJ...346..728J}; \citealt{1992ApJ...389..602J}; \citealt{2008ApJ...679..537F}). Various studies in molecular clouds using starlight polarization well support the RAT-A theory (for a review see \citealt{2015ARA&A..53..501A}). However, the test of RAT-A mechanism in star-forming regions using thermal dust polarization observation is only conducted recently (see e.g, \citealt{2022FrASS...9.3927T}; \citealt{2023ApJ...953...66N}; \citealt{2024ApJ...974..118N}; \citealt{2025arXiv250111634P}) with the advancement in the far infrared and sub-millimeter/millimeter polarimetric instruments. 

In this paper, we aim to study the grain alignment mechanism in the filamentary cloud regions of the reflection nebula, the Cocoon Nebula located at the east of the star-forming molecular cloud IC 5146 in the constellation of Cygnus using thermal dust polarization observations from JCMT/POL-2 at 850 $\mu$m. This nebula consists of three filaments namely F13, F13S and F13W \citep{2021ApJ...919....3C}. Dense cores named as C1, C2, C3 and C4 are identified in the F13 filament and dense cores C5, C6 and C7 in the F13S filament \citep{2024ApJ...970..122C}. These filaments and the core regions do not have bright embedded sources. A single B0 V star BD+46$^\circ$ 3474 (hereafter BD+46) is located at the center of this nebula (in the South of F13 filament) (e.g, \citealt{2008hsf1.book..108H}; \citealt{2014A&A...571A..93G}) surrounded by the HII region, Sharpless 125 (S125) \citep{1959ApJS....4..257S}. The distance of the Cocoon nebula is estimated to be 813 $\pm$ 106 pc \citep{2020ApJ...888...13W} and that of BD+46 to be 800 $\pm$ 80 pc \citep{2014A&A...571A..93G}. These filamentary regions of the nebula show significant variations in the gas column density and the dust temperature or equivalently the radiation field strength across the filaments from the outer regions towards the inner regions. These makes them ideal regions for the study of grain alignment mechanism in the context of RAT theory. In this work, we will study the grain alignment only in the F13 and F13S filaments. The rest of the paper is organized as follows: Section \ref{section:observations} provides the details of observational data, Section \ref{section:Analysis and Results} presents the data analysis and the results, Section \ref{section:Discussions} presents the discussions on the results and Section \ref{section:Conclusions} provides the conclusions. 

\section{Observations} \label{section:observations}
\subsection{Archival polarization data} \label{subsection:Polarization data}
In this work, we use the archival dust polarization data from \cite{2024ApJ...970..122C}. The polarimetric observations were done with the POL-2 instrument \citep{2016SPIE.9914E..03F} on the James Clerk Maxwell Telescope (JCMT) at 850 $\mu$m between 2021 June 9 and 2022 August 9 toward F13 region of Cocoon Nebula using the standard SCUBA-2/POL-2 daisy mapping mode that covers a total observing area of diameter $\approx$ $11'$ with a central $3'$ diameter area having the best sensitivity. The JCMT/POL-2 instrument has a Full Width at Half Maximum (FWHM) of $14''.1$ at 850 $\mu$m that corresponds to $\approx$ 0.056pc considering a distance of 813 pc of the Cocoon Nebula. For the reduction process, the data were reduced using the $pol2map$ routine in the Sub-Millimeter User Reduction Facility (SMURF) package of the Starlink Software. For the details on the polarimetric observations and the data reduction, please refer to Section 2.1 in \cite{2024ApJ...970..122C}. In our analysis, we use debiased polarization data with the selection criteria $I/\delta I > 10$ and $S/N > 2$ ($P/\sigma_P > 2$) due to limited and small sample of the polarization data. However, to see the validity of considering the data points with $2 < S/N < 3$, we check whether the inclusion of these data points with $2 < S/N < 3$ affects the trends in our analyses as described in Section \ref{subsubsection: Variation of P with I, N(H2) and Td}. In our analyses, we do weighted fitting in each analysis so that it will give more weight to the data with a higher signal-to-noise ratio.

\subsection{$H_2$ volume density and dust temperatures}
In this work, we use the maps of $\mathrm{H_2}$ column density and dust temperature of the IC 5146 Cocoon Nebula region obtained from the Herschel Gould Belt Survey (\citealt{2010A&A...518L.102A}; \citealt{2011A&A...529L...6A}). The $N(\mathrm{H_2})$ map is the one shown in \cite{2024ApJ...970..122C} with a resolution of $18''.2$. The $18''.2$ resolution of $N(\mathrm{H_2})$ is derived using the procedure described in \cite{2013A&A...550A..38P}. The dust temperature map has a resolution of $35''$. The maps were obtained using Herschel PACS/SPIRE data at 70, 160, 250, 350 and 500 $\mu$m. The maps are reprojected on the same grid as JCMT 850 $\mu$m intensity map and they have the same pixel scale of $4''$/pixel. For a particular pixel coordinate where polarization is detected, we create a circular region around this pixel coordinate in the $N(\mathrm{H_2})$ and $T_\mathrm{d}$ maps with a diameter of the JCMT beam size, $14''.1$ and take the mean value of all the pixel values found within this circular region. This mean value is used to compare with the JCMT/POL-2 data for that particular pixel coordinate.   

\begin{figure*}
    \centering
    \begin{tabular}{cc}
        \includegraphics[scale=0.43]{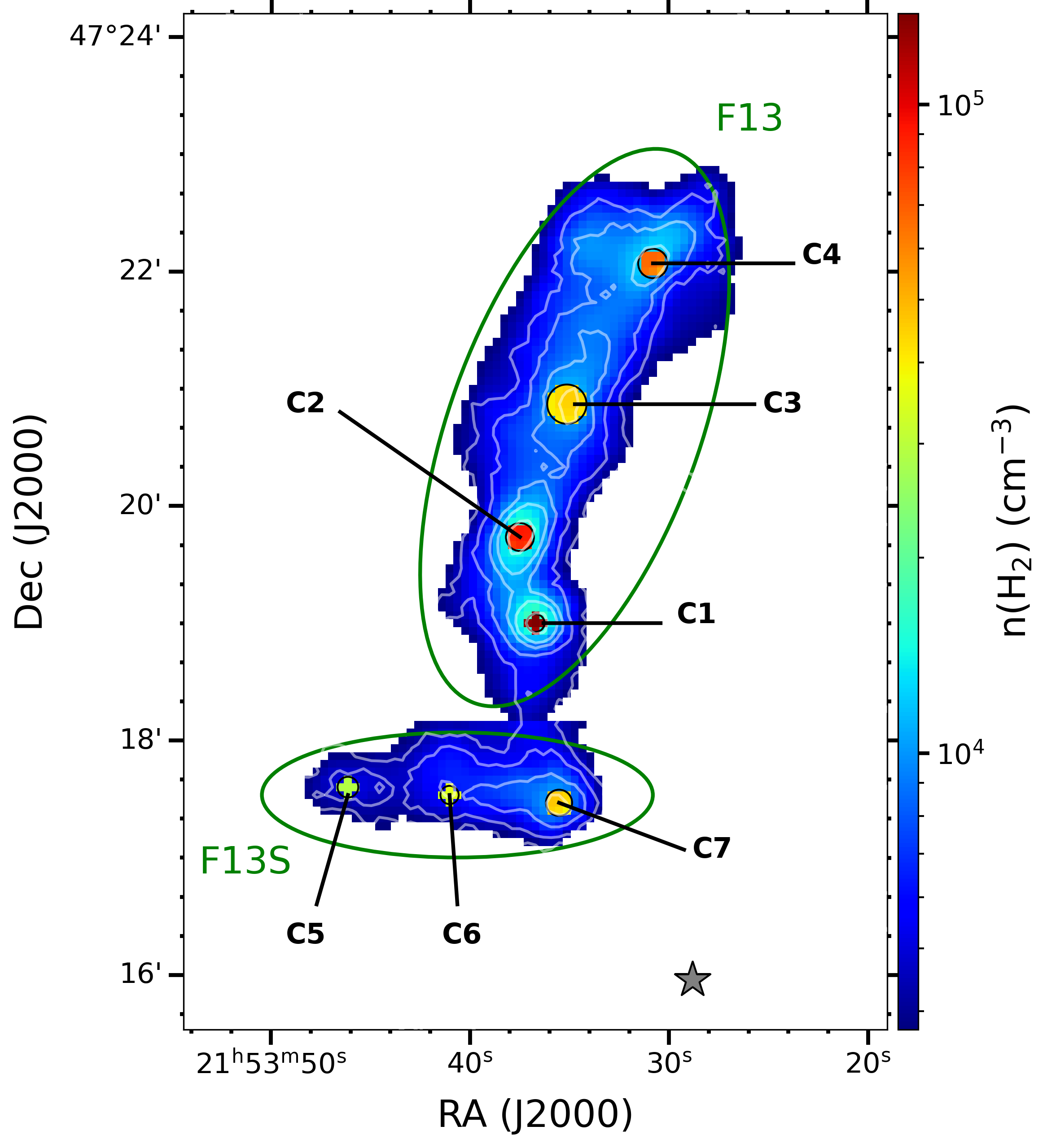} & 
        \hspace{5pt}
        \includegraphics[scale=0.43]{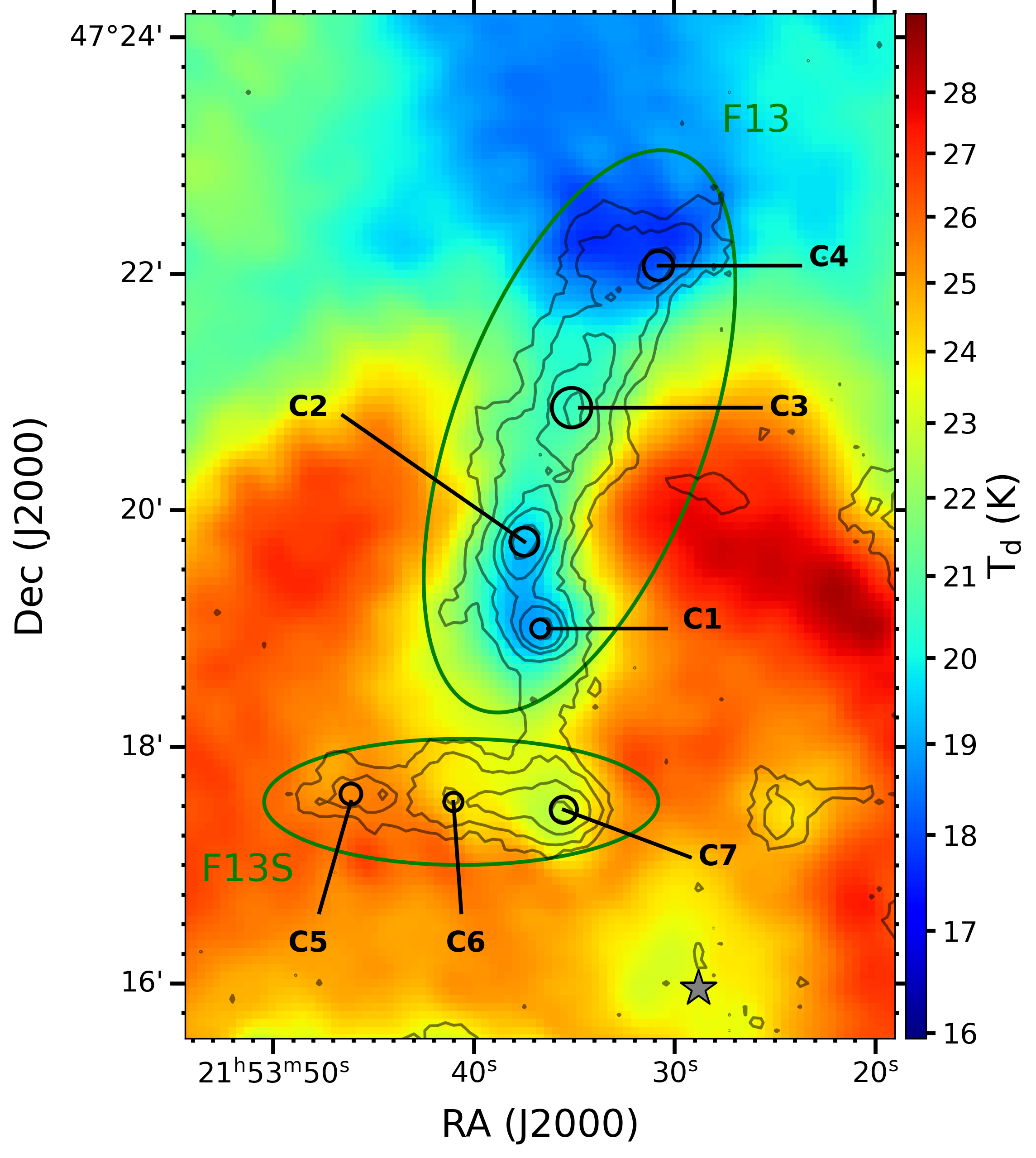} 
    \end{tabular}
    \caption{Maps of $\mathrm{H_{2}}$ volume density (left) and dust temperature (right) for the F13 and F13S filaments of the Cocoon nebula. The locations of the cores C1, C2, C3, C4 in the F13 filament and C5, C6 and C7 in the F13S filament as identified by \cite{2024ApJ...970..122C} are indicated. The contours are drawn at total intensity $I$ values of 12, 40, 80, 120 and 200 mJy/beam. The star symbol denotes the location of BD+46$^\circ$ 3474 (BD+46).}
    \label{Figure:Volume_density_Td_map}
\end{figure*}

\setlength{\tabcolsep}{0.3cm} 
\renewcommand{\arraystretch}{1}

\begin{table*}[ht]
    \centering
        \caption{Properties of the dense cores identified on the F13 and F13S filaments using the same procedure as described in \cite{2024ApJ...970..122C}}
    \begin{tabular}{ccccccccc}
        \hline
        Core ID & RA (J2000) & Dec (J2000) & Size 1 & Size 2 & Effective size & Size 1 & Size 2 & Effective size \\
           & ($^\circ$) & ($^\circ$) & ($''$) & ($''$) & ($''$) & (pc) & (pc) & (pc) \\ \hline
        C1 & 328.403 & 47.317 & 9.66 & 8.6 & 9.12 & 0.04 & 0.03 & 0.04 \\ \hline
        C2 & 328.406 & 47.329 & 15.54 & 13.15 & 14.30 & 0.06 & 0.05 & 0.06 \\ \hline
        C3 & 328.396 & 47.348 & 22 & 18.35 & 20.09 & 0.09 & 0.07 & 0.08 \\ \hline
        C4 & 328.378 & 47.368 & 15.63 & 14.45 & 15.03 & 0.06 & 0.06 & 0.06 \\ \hline
        C5 & 328.442 & 47.293 & 13.53 & 8.31 & 10.61 & 0.05 & 0.03 & 0.04 \\ \hline
        C6 & 328.421 & 47.292 & 10.96 & 7.95 & 9.34 & 0.04 & 0.03 & 0.04 \\ \hline
        C7 & 328.398 & 47.291 & 16.41 & 10.95 & 13.40 & 0.06 & 0.04 & 0.05 \\ \hline
    \end{tabular}
    
    \vspace{0.8em}
    
    \begin{minipage}{1\textwidth}
    Notes: (i) The given sizes 1 and 2 are the major and minor sizes and are deconvolved with the beam, i.e, $size = \sqrt{size^2_{uncorrected} - beam^2}$ \citep{2015A&C....10...22B}. \\
    (ii) The effective sizes both in $''$ and pc are the geometric means of Sizes 1 and 2, i.e, $Effective$ $size$ = $\sqrt{Size 1 \times Size 2}$. \\
    (iii) The physical sizes in pc are estimated considering a distance of 813 pc of the Cocoon Nebula.
    \end{minipage}
    \label{Table:Core_properties}
\end{table*}

We employ the column density map and estimate the $\mathrm{H_2}$ volume densities for the F13 and F13S filaments separately as they have different widths by assuming cylindrical geometries of these filaments so that we can consider their depths or thickness to be equal to their diameters or widths. \cite{2024ApJ...970..122C} estimated the width of F13 filament to be $\approx$ 0.32 pc and that of F13S to be $\approx$ 0.23 pc. However, we assume the embedded cores inside the filaments to have spherical geometries so that the core sizes or diameters can be considered equal to the depth of the cores. The properties of the cores identified using the same procedure as described in \cite{2024ApJ...970..122C} with FellWalker algorithm are given in Table \ref{Table:Core_properties}. For the derivation of the gas volume density, we consider the F13 and F13S filaments as two separate cylinders with diameters 0.32 pc and 0.23 pc, respectively and the embedded cores C1, C2, C3, C4, C5, C6 and C7 are assumed to have spherical geometries with effective angular diameters of $9''.12$, $14''.30$, $20''.09$, $15''.03$, $10''.61$, $9''.34$ and $13''.40$ that correspond to physical diameters of 0.04 pc, 0.06 pc, 0.08 pc, 0.06 pc, 0.04 pc, 0.04 pc and 0.05 pc (see Table \ref{Table:Core_properties}), respectively considering a distance of 813 pc for the Cocoon Nebula. The volume densities are derived using the following equation

\begin{equation}
{
n(\mathrm{H_2}) = \frac{N(\mathrm{H_2})}{d},
}
\end{equation}
where $d$ is the depth of the filaments. We also note that the estimation of width or depth of filaments is biased based on the angular resolutions of the observations (\citealt{2022A&A...657L..13P}). Also, our estimation of the volume density is based on the assumption of cylindrical geometries of these filaments and the spherical geometries of the embedded cores which may not reflect the true three-dimensional structure of the filaments. The exact three-dimensional geometry of star-forming filament is not yet fully known. However, a cylindrical geometry model of elongated filaments and spherical geometry model of the embedded cores stand to be a good approximation as per various studies (e.g., \citealt{2017ApJ...838...10M}; \citealt{2023ApJ...948..109K}). Our study which uses gas volume density values in this work is based on this assumption.

\begin{figure*}
    \centering
        \includegraphics[scale=0.75]{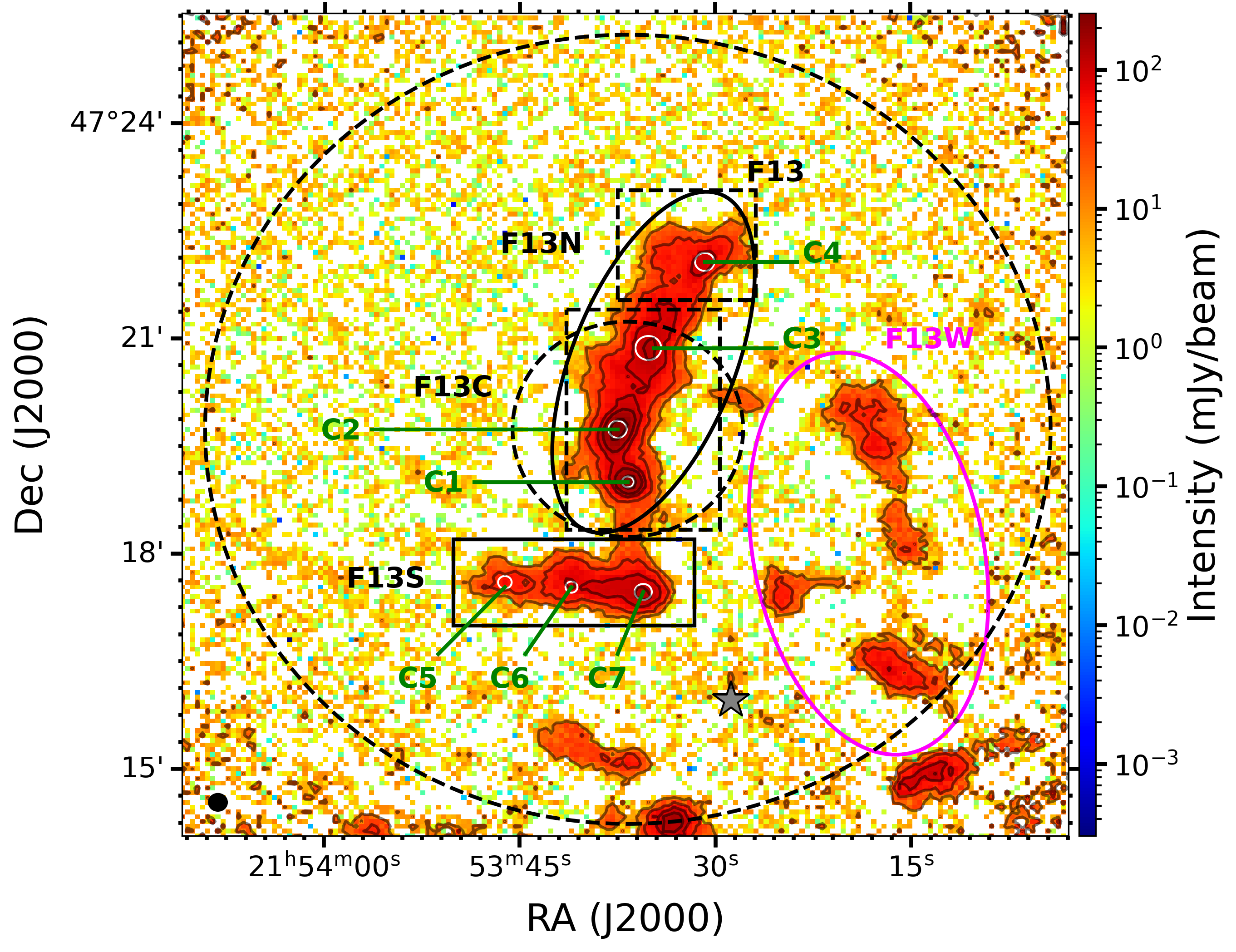} 
    \caption{Total emission intensity ($I$) map of the F13, F13S and F13W filaments of the Cocoon nebula observed by JCMT/POL-2 at 850$\mu$m with overlaid contours drawn at $I$ values of 12, 40, 80, 120 and 200 mJy/beam. The positions of the dense cores C1, C2, C3, C4 in the F13 filament and C5, C6 and C7 in the F13S filaments are indicated with white circles and labelled with green colors. The dashed small and large black circles are drawn at diameters of $3'$ and $11'$, which show the central area of best sensitivity and the total observing area, respectively. The solid black and magenta edge-colored ellipses represent the regions of the F13 which we study and F13W filaments and the solid black edge-colored rectangle represents the F13S filament region which we study. The F13 filament is further classified into two sub-regions shown with two dashed black rectangles and named as F13N and F13C that denote the North and the Center regions of this filament. Also, the beam size, $14''.1$ of JCMT/POL-2 at 850$\mu$m is indicated with a solid black circle. The star symbol denotes the location of BD+46$^\circ$ 3474 (BD+46).}
    \label{Figure:Intensity_map}
\end{figure*}

We combine the maps of $n(\mathrm{H_2})$ for both the F13 and F13S filaments into a single map and is shown in Figure \ref{Figure:Volume_density_Td_map} (left). The F13 filament has comparatively higher volume densities than the F13S filament with the highest values of more than $10^4$ $\mathrm{cm^{-3}}$ being found in the cores C1 and C2 in F13. For F13S, the core C7 has high volume density of the order of $\approx$ $10^4$ $\mathrm{cm^{-3}}$. From the dust temperature map shown in Figure \ref{Figure:Volume_density_Td_map} (right), we find that the core regions of C1, C2 and C4 in F13 show significant reduction in $T_\mathrm{d}$ values up to around 17 K. The F13 filament has a range of $T_\mathrm{d}$ values from $17.5-23$ K and the F13S filament which is comparatively less denser than F13 has a range of $T_\mathrm{d}$ values from $23-26$ K. This F13S filament is nearby to the illuminating star BD+46 shown with a star symbol in the map. The higher $T_\mathrm{d}$ values in F13S can be due to its close proximity to BD+46 and being less denser. We find that the dust temperature decreases from outside towards the inner regions of both the filaments. The dust temperatures in these filaments can be considered due to both the diffuse interstellar radiation field (ISRF) and the strong radiation from BD+46.  

% \begin{figure*}
%     \centering
%         \includegraphics[scale=0.75]{Cocoon_Intensity_map_Final_1.png} 
%     \caption{Total emission intensity ($I$) map of the F13, F13S and F13W filaments of the Cocoon nebula observed by JCMT/POL-2 at 850$\mu$m with overlaid contours drawn at $I$ values of 12, 40, 80, 120 and 200 mJy/beam. The positions of the dense cores C1, C2, C3, C4 in the F13 filament and C5, C6 and C7 in the F13S filaments are indicated with white circles and labelled with green colors. The dashed small and large black circles are drawn at diameters of $3'$ and $11'$ that shows the central area of best sensitivity and the total observing area, respectively. The solid black and magenta edgecolored ellipses represent the regions of the F13 which we study and F13W filaments and the solid black edgecolored rectangle represents the F13S filament region which we study. The F13 filament is further classified into two sub-regions shown with two dashed black rectangles and named as F13N and F13C that denote the North and the Center regions of this filament. Also, the beam size, $14''.1$ of JCMT/POL-2 at 850$\mu$m is indicated with a solid black circle. The star symbol denotes the location of BD+46$^\circ$ 3474 (BD+46).}
%     \label{Figure:Intensity_map}
% \end{figure*}

\section{Analysis and Results} \label{section:Analysis and Results}
\subsection{Data Analyses}
\subsubsection{Total emission intensity and polarization vector maps}
Figure \ref{Figure:Intensity_map} shows the map of the total emission intensity $I$ of the filamentary regions of the Cocoon Nebula observed by JCMT/POL-2 at 850 $\mu$m. The positions of the dense cores C1, C2, C3, C4 in the F13 filament and C5, C6, C7 in the F13S filament as identified by \cite{2024ApJ...970..122C} are indicated in the figure with black line circles. The small and the large dashed black line circles denote the central region of $3'$ in diameter having the best sensitivity and a total observing area of $11'$ in diameter, respectively. We marked two regions as F13N and F13C that denote the North and the Center regions within the F13 filament with dashed black rectangles and the F13S filament with a solid line black rectangle. Hereafter, we will use the color notation of green for F13N, magenta for F13C and black for F13S in all the analyses. We find that the regions of the cores C1 and C2 have higher intensities than other core regions.

% \begin{figure*}
%     \centering
%         \includegraphics[scale=0.75]{Cocoon_Intensity_map_Final_1.png} 
%     \caption{Total emission intensity ($I$) map of the F13, F13S and F13W filaments of the Cocoon nebula observed by JCMT/POL-2 at 850$\mu$m with overlaid contours drawn at $I$ values of 12, 40, 80, 120 and 200 mJy/beam. The positions of the dense cores C1, C2, C3, C4 in the F13 filament and C5, C6 and C7 in the F13S filaments are indicated with white circles and labelled with green colors. The dashed small and large black circles are drawn at diameters of $3'$ and $11'$ that shows the central area of best sensitivity and the total observing area, respectively. The solid black and magenta edgecolored ellipses represent the regions of the F13 which we study and F13W filaments and the solid black edgecolored rectangle represents the F13S filament region which we study. The F13 filament is further classified into two sub-regions shown with two dashed black rectangles and named as F13N and F13C that denote the North and the Center regions of this filament. Also, the beam size, $14''.1$ of JCMT/POL-2 at 850$\mu$m is indicated with a solid black circle. The star symbol denotes the location of BD+46$^\circ$ 3474 (BD+46).}
%     \label{Figure:Intensity_map}
% \end{figure*}
We see that overall the core regions show comparatively higher intensity values than other regions. The intensity is found to be small in the outer regions of the filaments and it increases inwards.

\begin{figure}
    \centering
        \includegraphics[scale=0.5]{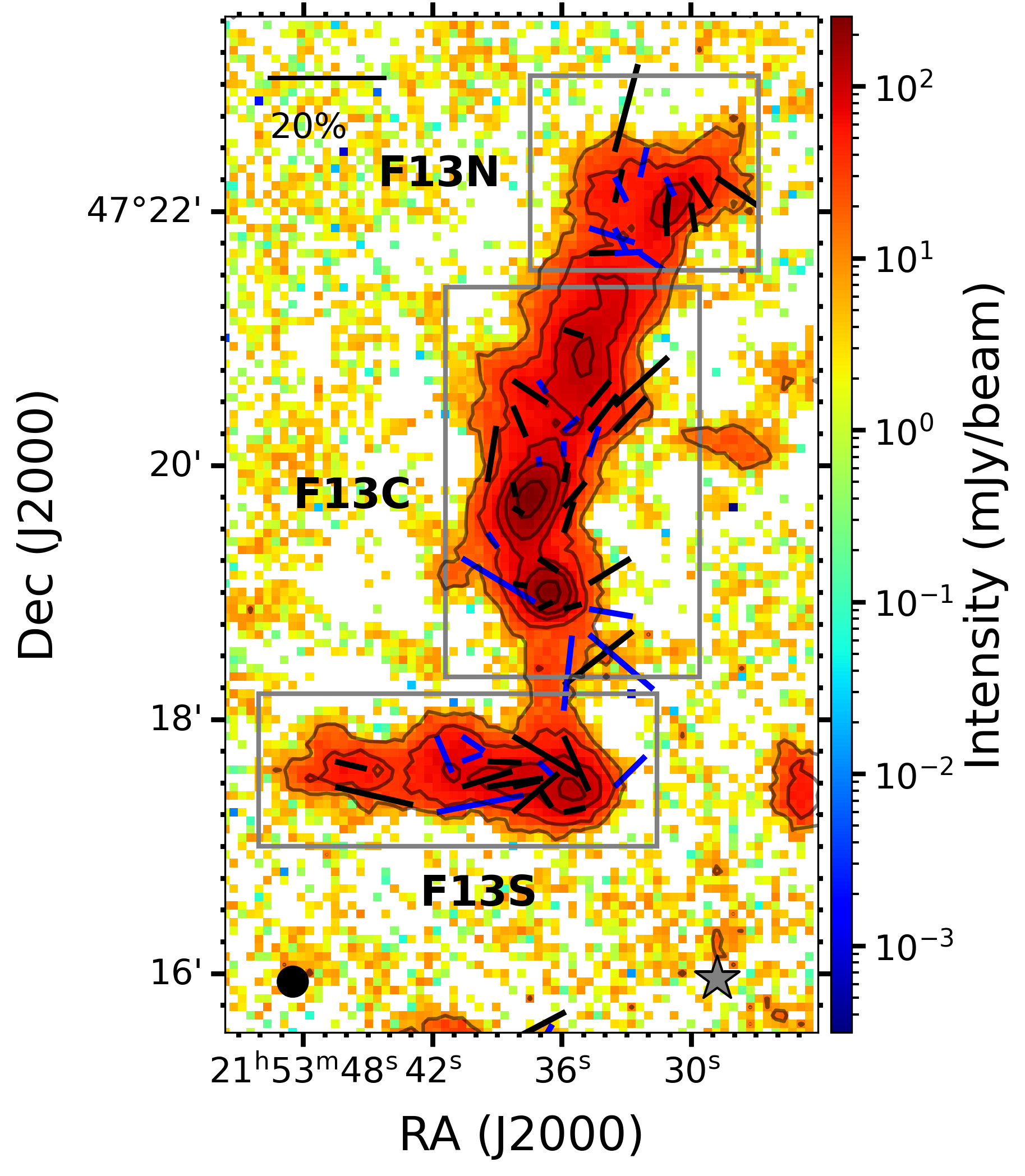} 
    \caption{Same map as in Figure \ref{Figure:Intensity_map}, but overlaid with the polarization vectors. The blue vectors are for $2 < P/\sigma_P < 3$ and the black vectors are for $P/\sigma_P \geq 3$. The length of the vectors are proportional to the polarization fraction $P$ and their orientations determine the magnetic field orientations. A reference length scale of $P$ is also indicated.} 
    \label{Figure:Intensity_P_map}
\end{figure}

Figure \ref{Figure:Intensity_P_map} shows the distribution of the polarization vectors overplotted on the total intensity map. The lengths of the vectors are proportional to the polarization fraction $P$ and their orientations determine the magnetic field orientations. A reference scale length of $P$ is indicated in the Figure. We see that the value of $P$ is large in the outer regions having small intensities and decreases in the inner regions associated with higher intensities. 

\begin{figure*}
    \centering
        \includegraphics[scale=0.56]{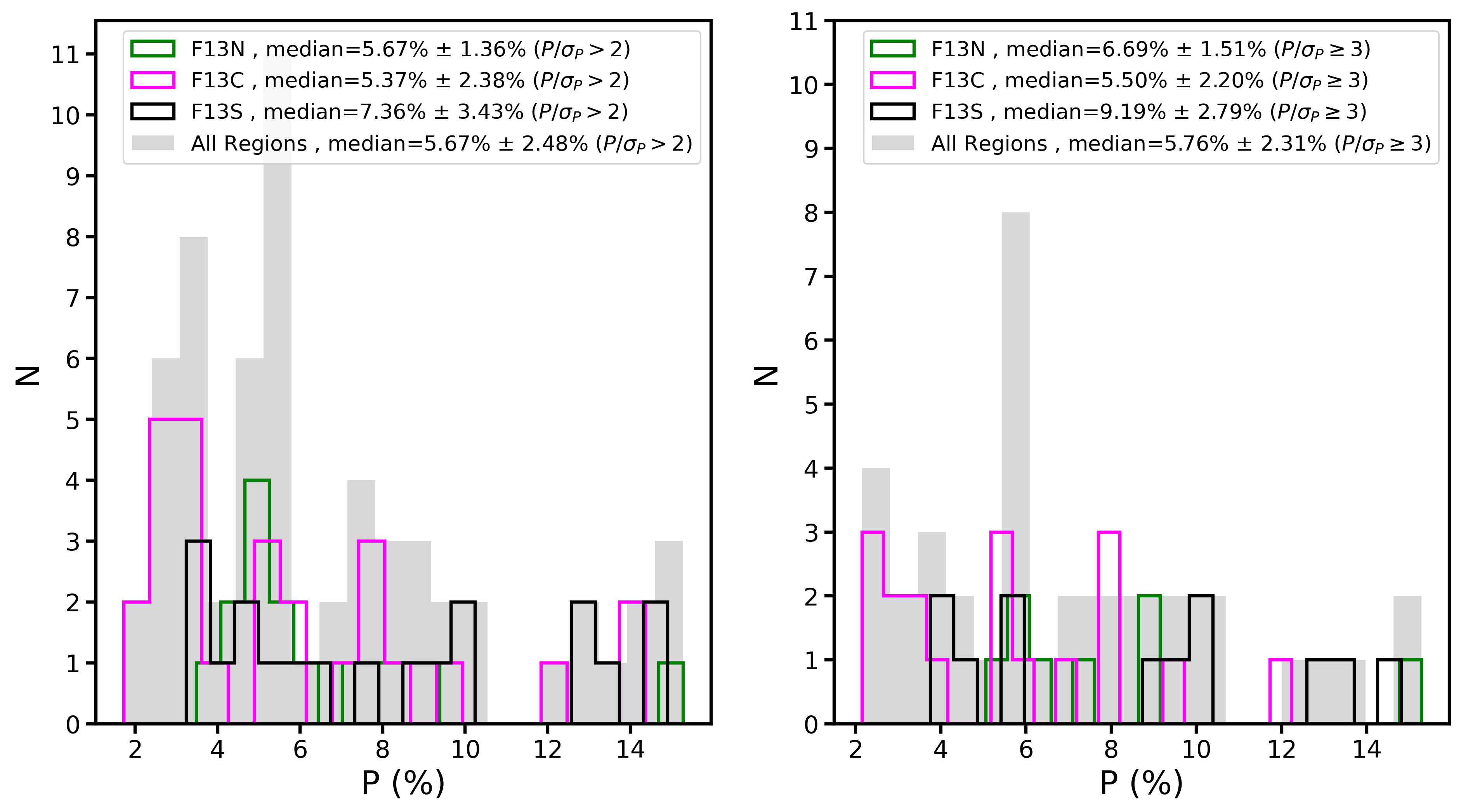} 
    \caption{Histograms of the polarization fraction $P$ for the F13N, F13C, F13S regions and the combined regions of the F13 and F13S filaments for $P/\sigma_P > 2$ (left) and $P/\sigma_P \geq 3$ (right). The solid green, magenta and black lines are for the F13N, F13C and F13S regions respectively. The filled gray color is for the combined regions of F13 and F13S filaments.}
    \label{Figure:Histogram_P}
\end{figure*}

The left and right panels in Figure \ref{Figure:Histogram_P} show the histograms of $P$ for the F13N, F13C, F13S and overall region of both the filaments for $S/N > 2$ and $S/N \geq 3$, respectively. The filled gray color is for all the regions. The median polarization fractions for the F13N, F13C and F13S regions with $S/N > 2$ are $5.67 \pm 1.36$\%, $5.37 \pm 2.38$\%, $7.36 \pm 3.43$\% and for $S/N > 3$ are $6.69 \pm 1.51$\%, $5.50 \pm 2.20$\%, $9.19 \pm 2.79$\%, respectively. The median polarization fractions for all the regions with $S/N > 2$ and $S/N > 3$ are $5.67 \pm 2.48$\% and $5.76 \pm 2.31$\%, respectively. Both the histograms show nearly similar distributions of $P$ and the median values for both $S/N > 2$ and $S/N > 3$ are nearly similar. The F13S filament is found to have higher $P$ values than the F13 filament. This F13S filament is comparatively less denser than the F13 filament and located more closer to the B-type star BD+46. The high radiations from this star may increase the grain alignment in the outer regions of the F13S filament producing higher polarization fraction.

\begin{figure}
    \centering
        \includegraphics[scale=0.52]{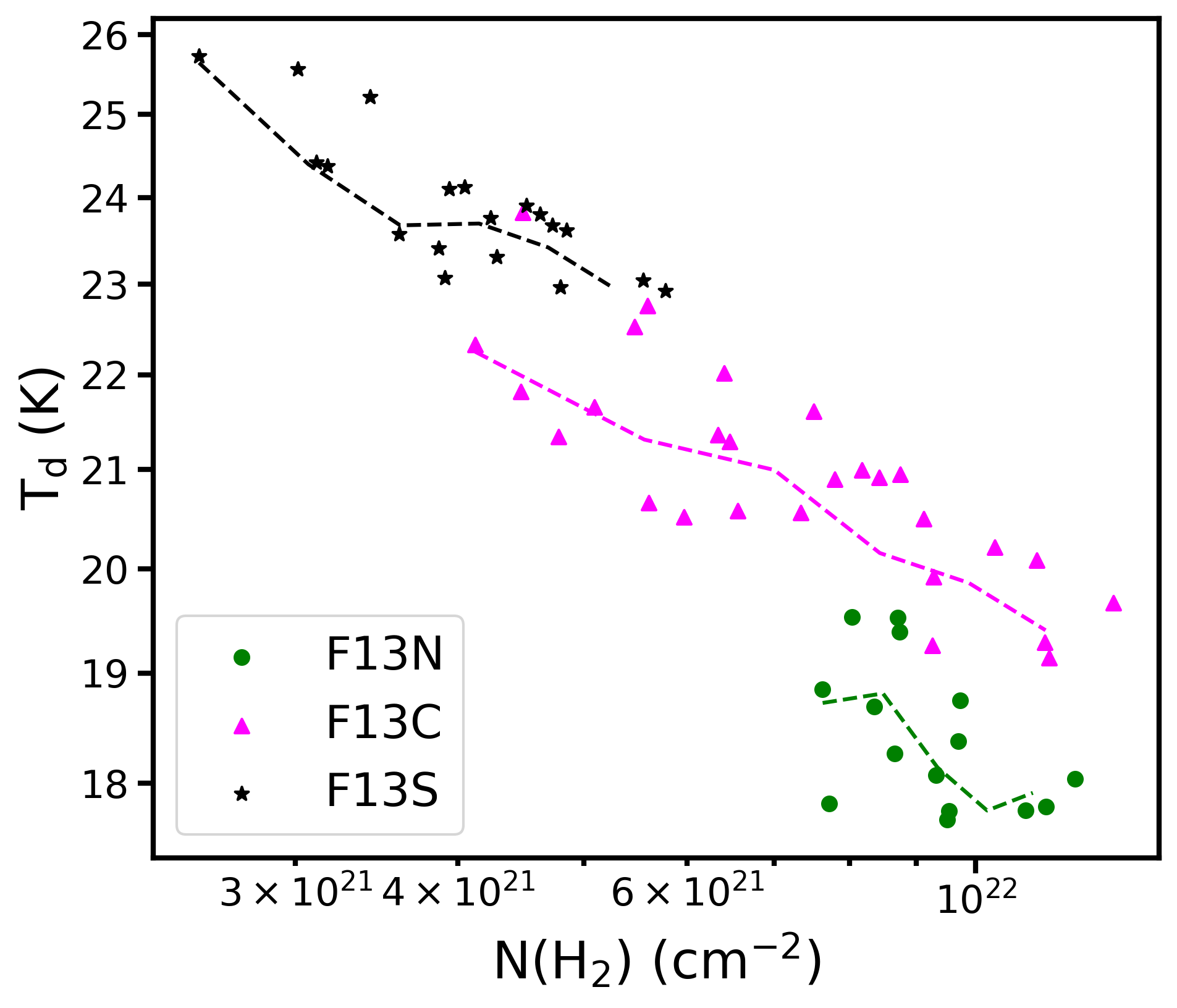} 
    \caption{Variation of dust temperature with gas column density in each region.}
    \label{Figure:Td_NH2}
\end{figure}

\begin{figure*}
    \centering
    \begin{tabular}{cccc}
        \includegraphics[scale=0.5]{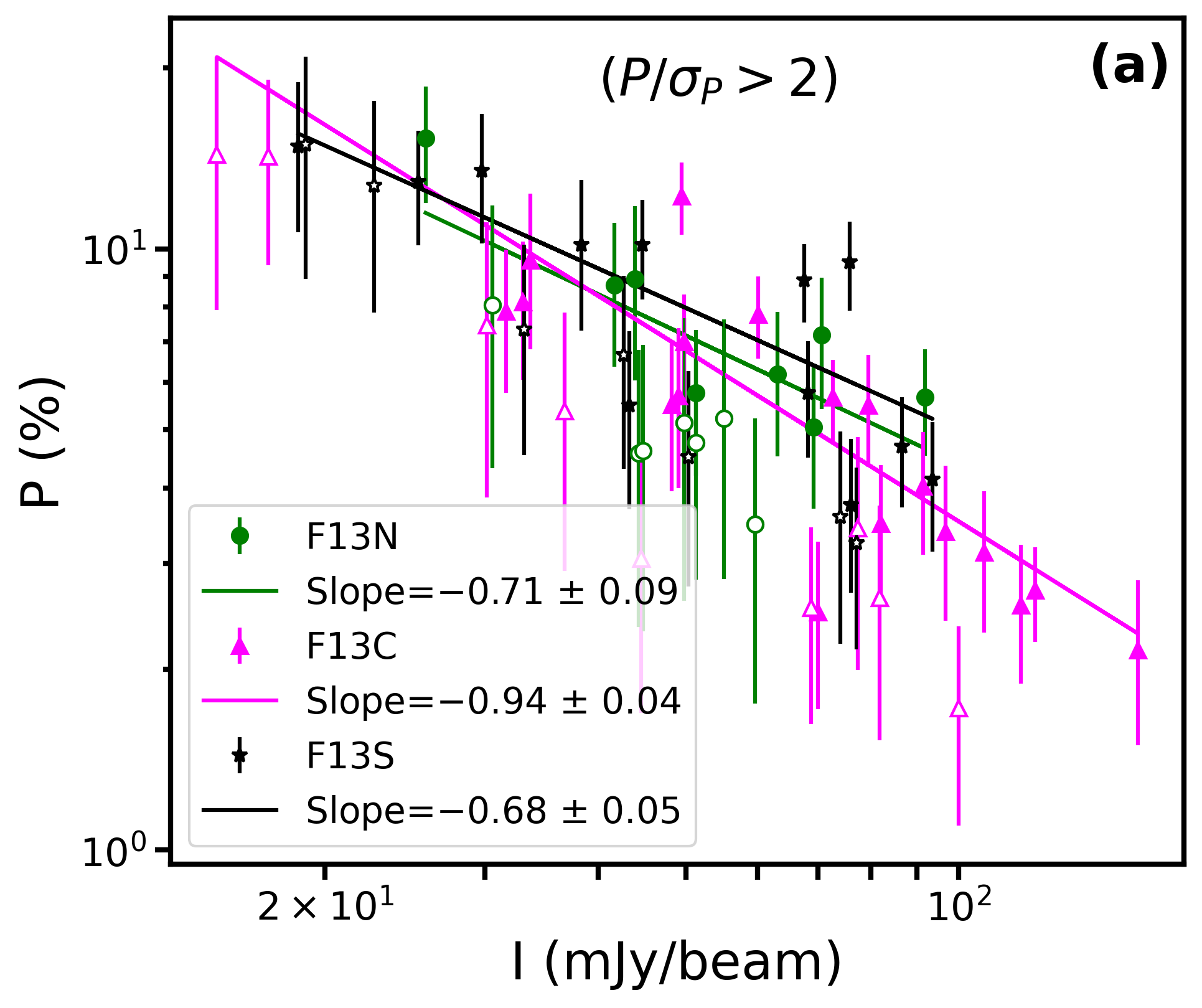} & 
        \hspace{5pt}
        \includegraphics[scale=0.5]{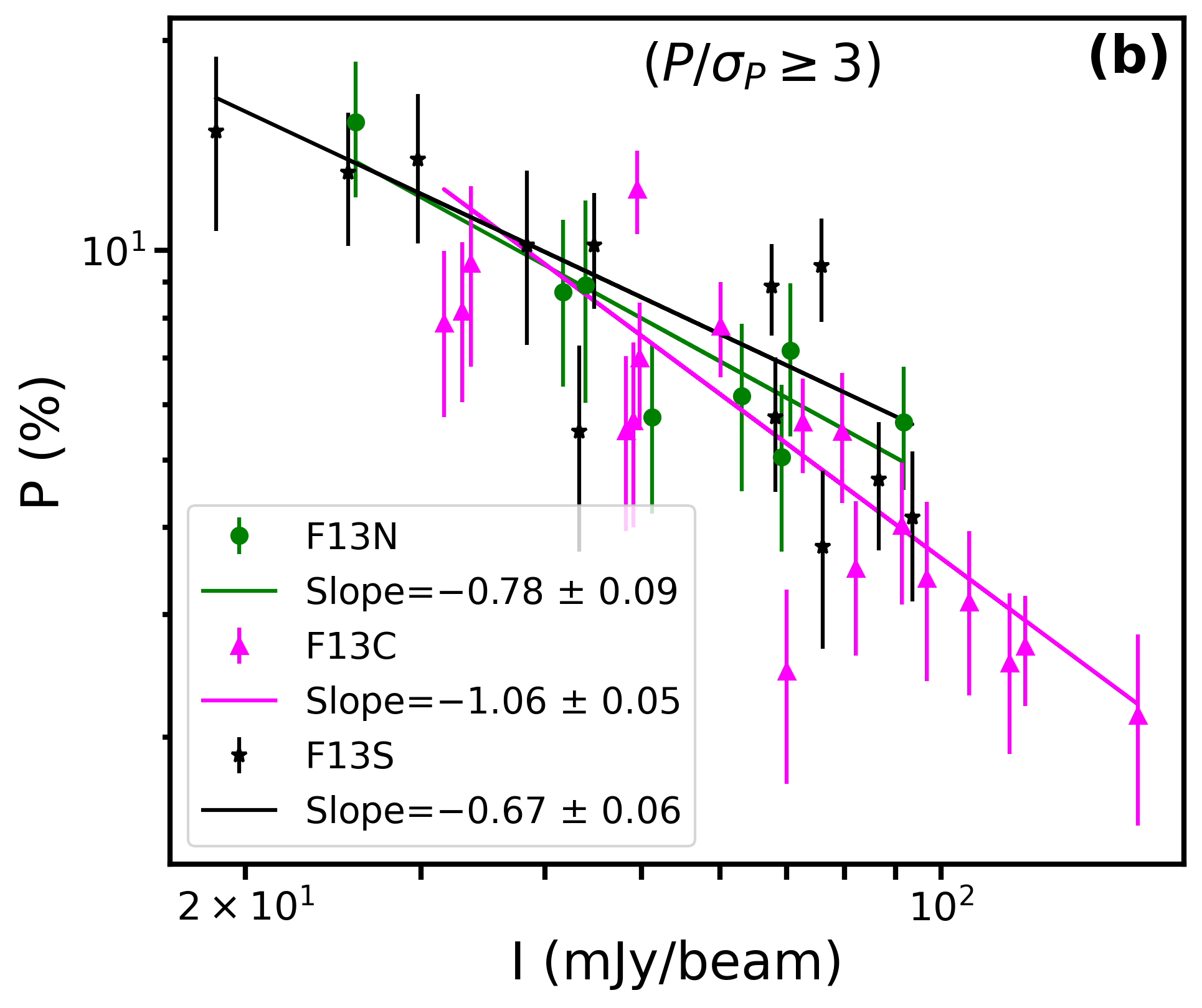} &  \\
        %\hspace{5pt}
        \includegraphics[scale=0.5]{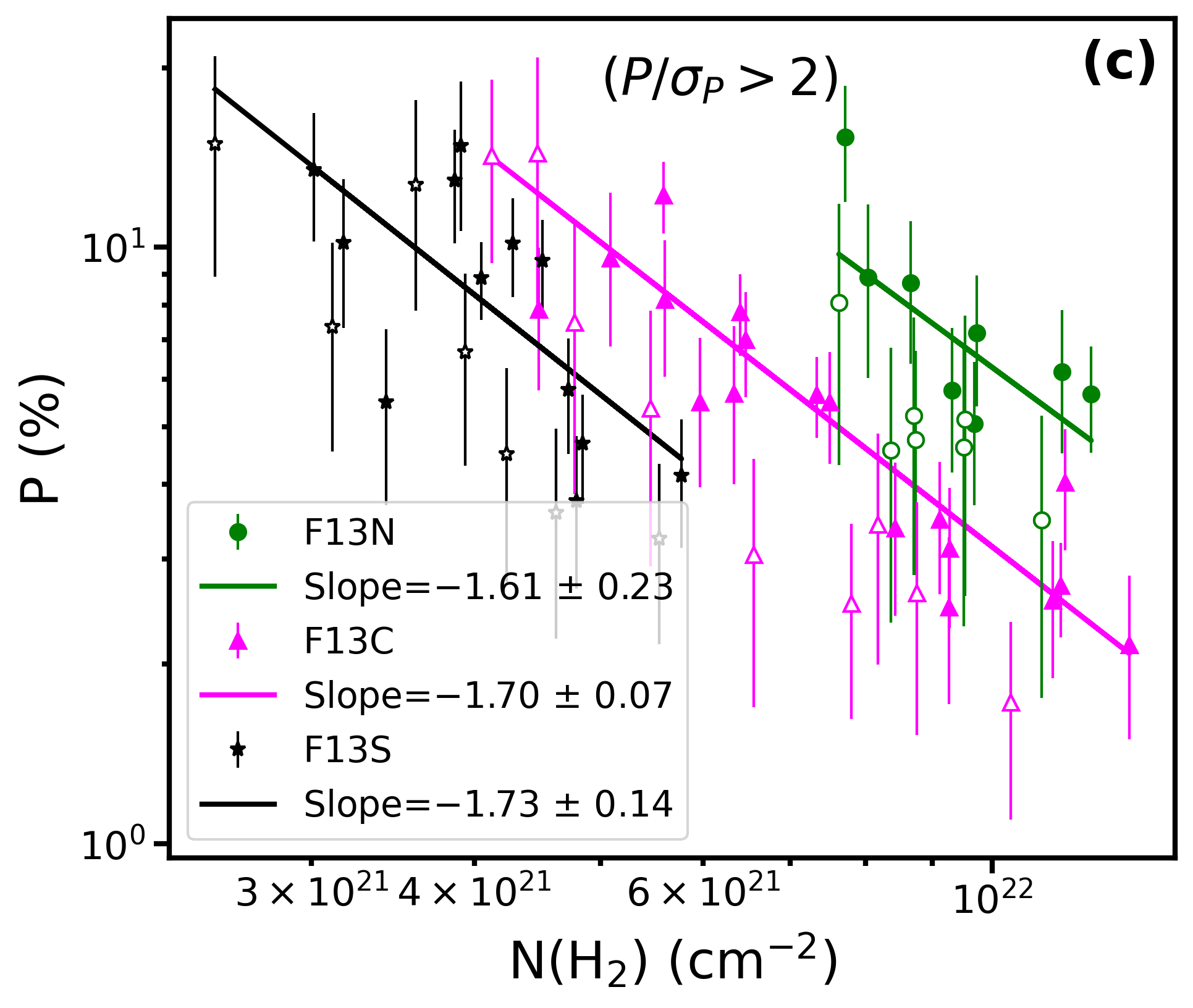} & 
        \hspace{5pt}
        \includegraphics[scale=0.5]{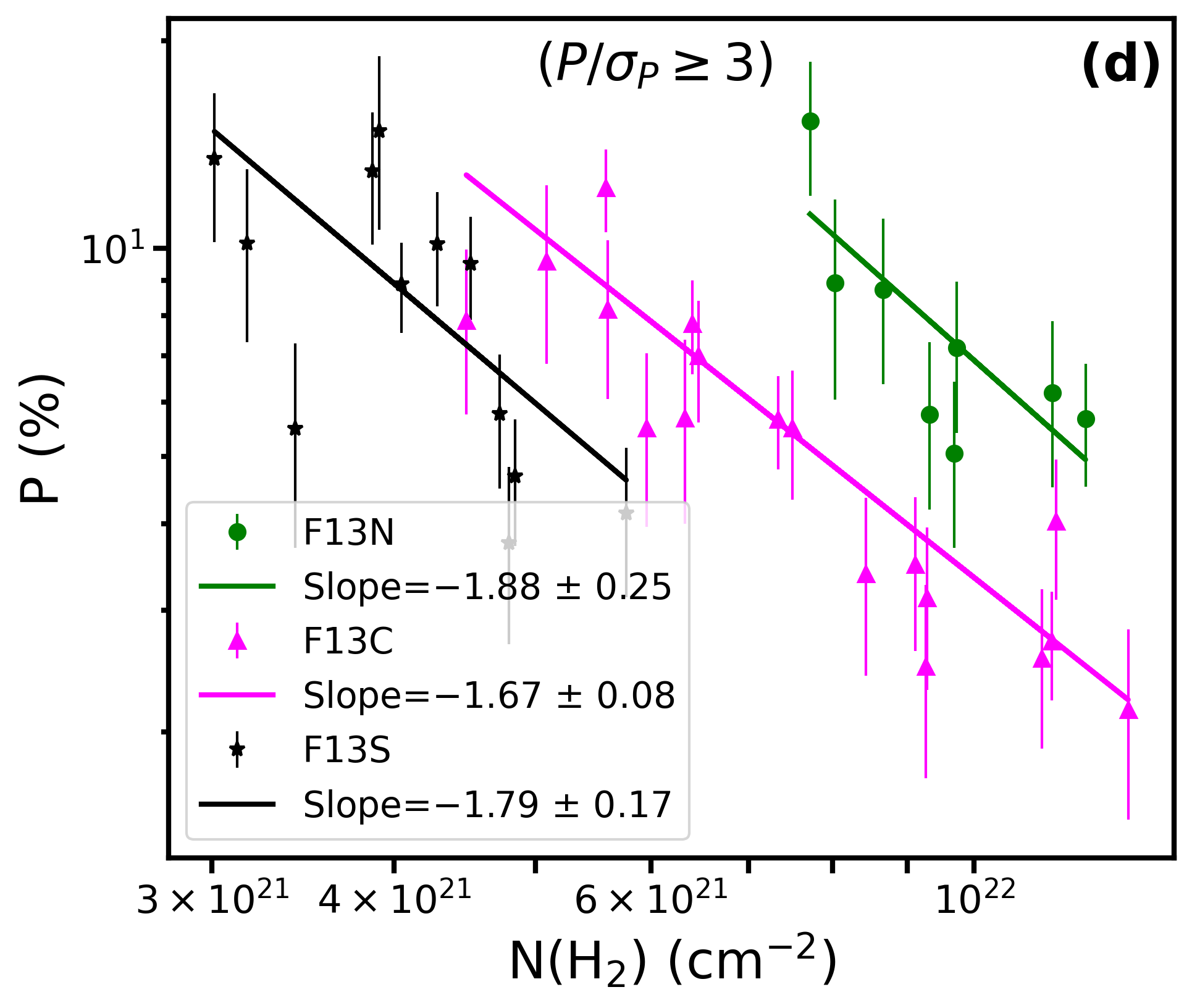}     
    \end{tabular}
    \caption{Variations of (a) $P$ with $I$ for $P/\sigma_P > 2$, (b) $P$ with $I$ for $P/\sigma_P \geq 3$, (c) $P$ with $N(\mathrm{H_2})$ for $P/\sigma_P > 2$, (d) $P$ with $N(\mathrm{H_2})$ for $P/\sigma_P \geq 3$. The white facecolor data points are the data points associated with $2 < P/\sigma_P < 3$. The solid lines are the weighted best power-law fits.}
    \label{Figure:P_I_NH2}
\end{figure*}

\begin{figure*}
    \centering
    \begin{tabular}{cc}
        \includegraphics[scale=0.45]{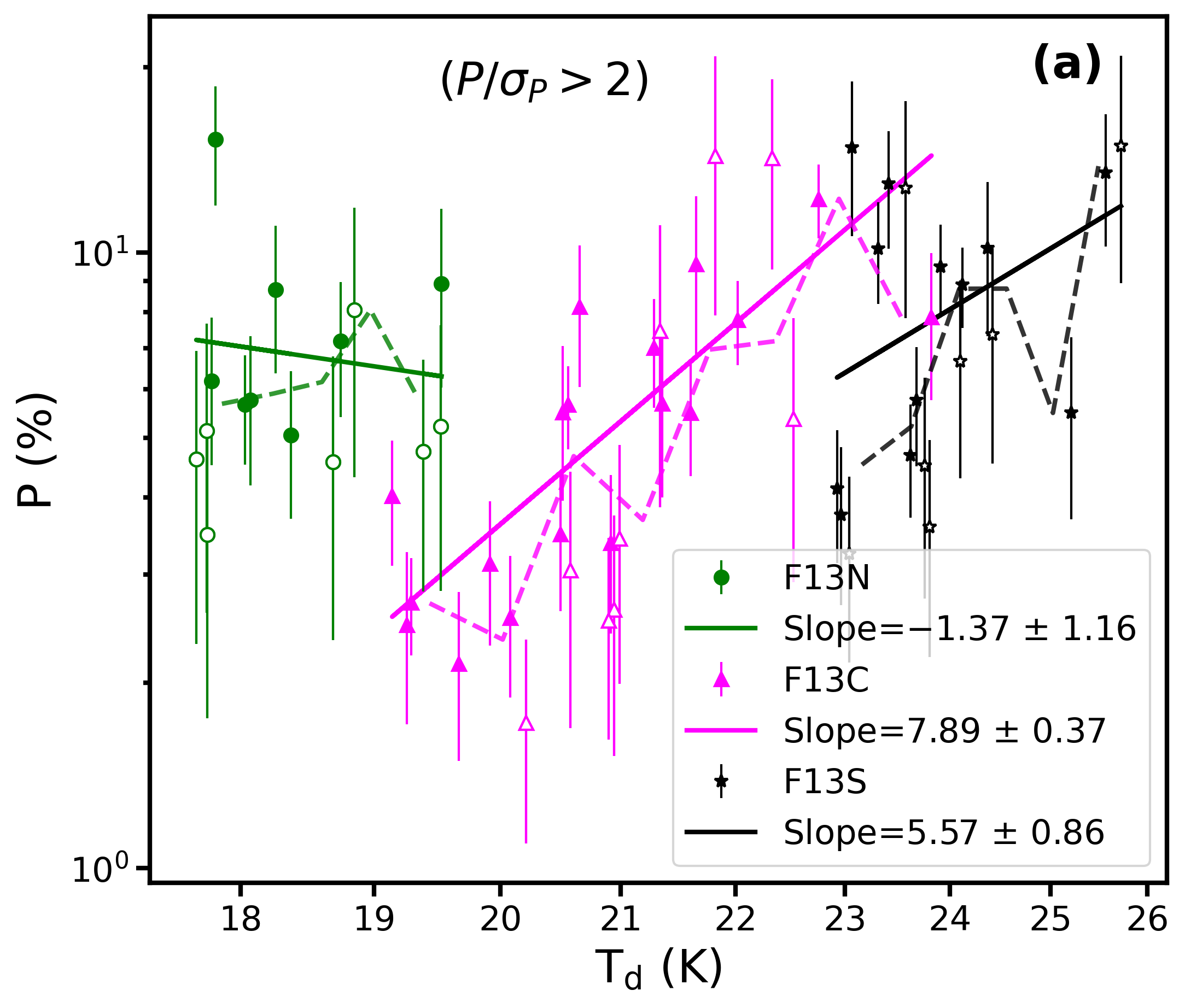} & 
        \hspace{5pt}
        \includegraphics[scale=0.45]{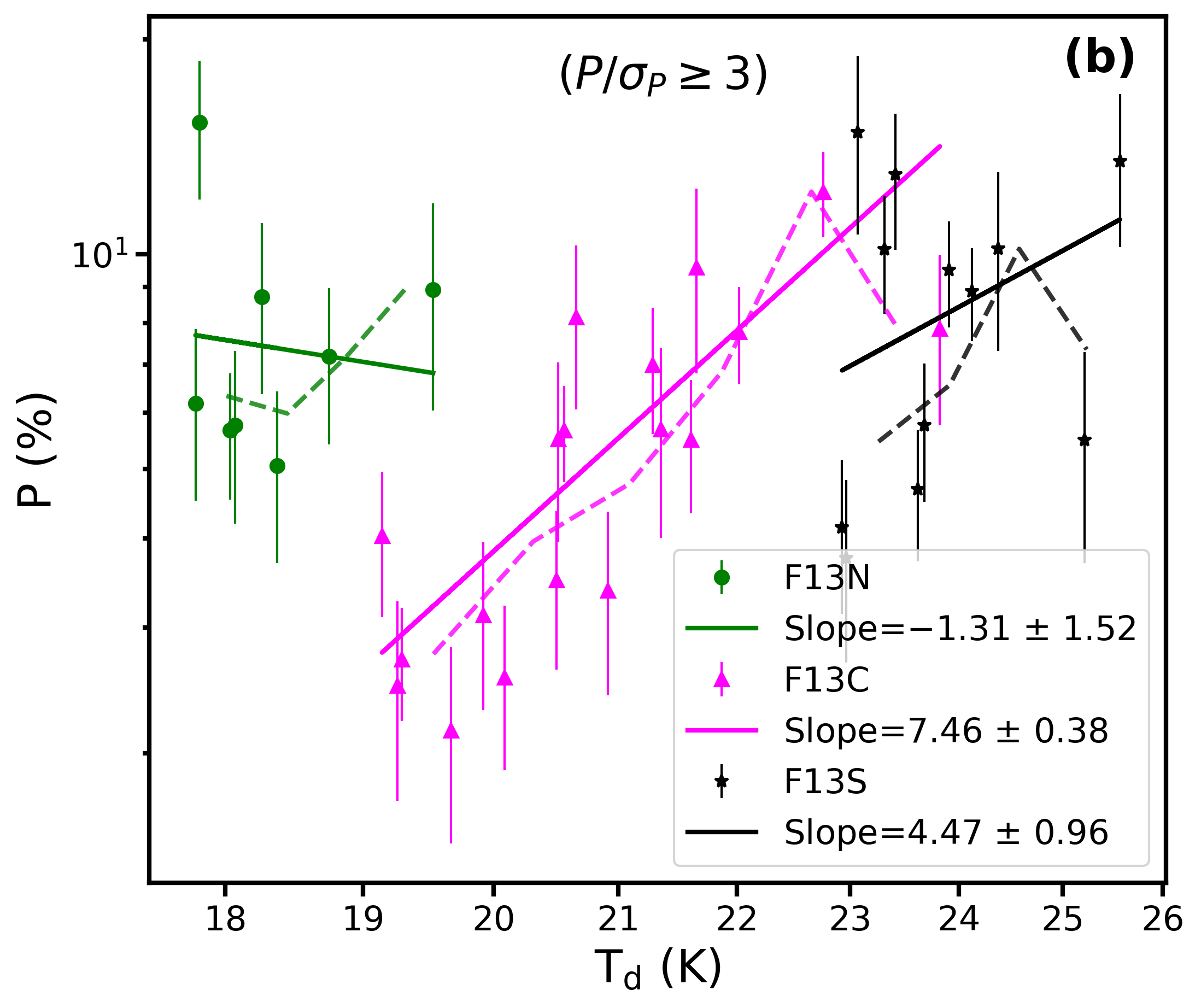} 
    \end{tabular}
    \caption{Variations of $P$ with $T_\mathrm{d}$ (a) for $P/\sigma_P > 2$ and (b) for $P/\sigma_P \geq 3$. The white facecolor data points are the data points associated with $2 < P/\sigma_P < 3$. The dashed lines are the weighted running means and the solid lines are the weighted best power-law fits.}
    \label{Figure:P_Td}
\end{figure*}

\subsubsection{Variation of polarization fraction with total intensity, gas column density and dust temperature} \label{subsubsection: Variation of P with I, N(H2) and Td}
In Figure \ref{Figure:Td_NH2}, the variation of $T_\mathrm{d}$ with $N(\mathrm{H_2})$ in each of the regions is shown. We see that the dust temperature decreases as the gas column density increases in each of the regions. This implies that denser regions are associated with lower dust temperatures or lower radiation field strength due to the absence of bright embedded sources. The dust temperatures are due to the diffuse ISRF and the radiation from the BD+46 star only. In these denser regions, because of less radiation field strength the grains may not be able to achieve suprathermal rotation and can be randomized, according to the RAT alignment theory \citep{2021ApJ...908..218H}, after gas-grain collisions. This would result in the decrease of $P$ in these denser regions. We analyse the variations of $P$ with $I$, $N(\mathrm{H_2})$ and $T_\mathrm{d}$ in this section.

Figure \ref{Figure:P_I_NH2} (a) and (b) show the variations of the polarization fraction $P$ with the increase in total intensity $I$ for $S/N > 2$ and $S/N \geq 3$, respectively. We see that $P$ is very high up to around 15\% in the outer regions of both the filaments with $I < 30$ mJy/beam and it decreases with the increase in $I$ towards the filament's spine. The variations of $P$ with $I$ are fitted with power-law best-fit lines of the form $P=k_1I^{a_1}$, where $k_1$ is a constant and $a_1$ is the slope of the fit. The best fits result in $a_1$ values of $-0.71 \pm 0.09$, $-0.94 \pm 0.04$ and $-0.68 \pm 0.05$ for the F13N, F13C and F13S filaments, respectively for $S/N > 2$ and $-0.78 \pm 0.09$, $-1.06 \pm 0.05$ and $-0.67 \pm 0.06$ for $S/N \geq 3$. The data points with white facecolors and with edgecolors of same color notation defined for the corresponding regions are for $2 < S/N < 3$. We find that the slope values of the weighted fits in both $S/N > 2$ and $S/N \geq 3$ are almost similar overall but slightly changes for the F13C region. The inclusion of $2 < S/N < 3$ data points does not significantly affect the observed trends found when we exclude them.

Figure \ref{Figure:P_I_NH2} (c) and (d) show the variations of $P$ with the increase in gas column density $N(\mathrm{H_2})$ for $S/N > 2$ and $S/N \geq 3$, respectively. We find that $P$ decreases with the increase in the values of $N(\mathrm{H_2})$ in each of the regions. Best power-law fits of the form $P=k_2[N(\mathrm{H_2})]^{a_2}$, where $k_2$ is a constant and $a_2$ is the slope, result in $a_2$ values of $-1.61 \pm 0.23$, $-1.70 \pm 0.07$ and $-1.73 \pm 0.14$ for the F13N, F13C and F13S filaments, respectively for $S/N > 2$ and $-1.88 \pm 0.25$, $-1.67 \pm 0.08$ and $-1.79 \pm 0.17$ for $S/N \geq 3$. Here also, inclusion of $2 < S/N < 3$ data points does not significantly affect the observed trends. We see that $P$ decreases with the increase in both the total intensity and the gas column density.

%\subsubsection{Variation of polarization fraction with dust temperature}
Figure \ref{Figure:P_Td} (a) and (b) show the variations of $P$ with the increase in dust temperatures $T_\mathrm{d}$ for each of the regions for $S/N > 2$ and $S/N \geq 3$, respectively. $P$ shows an overall increase with the increase in $T_\mathrm{d}$ in the F13C and F13S regions. The F13N shows some spread in the data, however the weighted running mean shows a tendency to slightly increase in $P$ with $T_\mathrm{d}$ for $S/N \geq 3$ and $P$ shows nearly flat distribution with $T_\mathrm{d}$ overall for $S/N > 2$ but we find some increase in $P$ also. We see some data points having smaller $P$ values at larger $T_\mathrm{d}$ values which may be due to local magnetic field tangling. Best power-law fits of the form $P=k_3T_\mathrm{d}^{a_3}$ with $k_3$ a constant and $a_3$ the slope results in $a_3$ values of $-1.37 \pm 1.16$, $7.89 \pm 0.37$ and $5.57 \pm 0.86$ for the F13N, F13C and F13S regions, respectively for $S/N > 2$ and $-1.31 \pm 1.52$, $7.46 \pm 0.38$ and $4.47 \pm 0.96$ for $S/N \geq 3$. The observed trends for both $S/N > 2$ and $S/N \geq 3$ are nearly similar.

From our analysis of variations of $P$ with $I$, $N(\mathrm{H_2})$ and $T_\mathrm{d}$ for both $S/N > 2$ and $S/N \geq 3$, because of the close similarity in the observed trends and also close similarity in the distributions of $P$ (see Figure \ref{Figure:Histogram_P}) for both of these signal-to-noise ratios, we consider $2 < S/N < 3$ data points in the subsequent analyses so as to increase the statistics of our small data sample. However, we perform weighted analysis considering the uncertainties.

The decrease in $P$ with $I$ and $N(\mathrm{H_2})$ as seen in Figure \ref{Figure:P_I_NH2} can be caused due to decrease in grain alignment efficiency in denser regions or magnetic field tangling or both. We need to find out whether magnetic field tangling has any significant contribution to causing the depolarization or not. For that, we analyse the effect of magnetic field tangling on the depolarization in the following section.

% \begin{figure*}
%     \centering
%         \includegraphics[scale=0.56]{Cocoon_Histogram_P_Final_1.png} 
%     \caption{Histograms of the polarization fraction $P$ for the F13N, F13C, F13S regions and the combined regions of the F13 and F13S filaments for $P/\sigma_P > 2$ (left) and $P/\sigma_P \geq 3$ (right). The solid green, magenta and black lines are for the F13N, F13C and F13S regions respectively. The filled gray color is for the combined regions of F13 and F13S filaments.}
%     \label{Figure:Histogram_P}
% \end{figure*}

% \begin{figure}
%     \centering
%         \includegraphics[scale=0.52]{Cocoon_Td_NH2_Final.png} 
%     \caption{Variation of dust temperature with gas column density in each region.}
%     \label{Figure:Td_NH2}
% \end{figure}

\subsubsection{Polarization angle dispersion function; effect of magnetic field tangling}
The decrease in the polarization fraction with the increase in total intensity and gas column density, widely known as polarization hole can be caused by the decrease in the alignment efficiency of the grains in the denser regions as expected by RAT-A theory or by fluctuations in the magnetic fields along the line-of-sight due to turbulence. To study whether the observed depolarization as the intensity and gas column density increases is due to the effect of magnetic field tangling along the line of sight or the decrease in grain alignment efficiency in the denser regions or both, it is important to disentangle the effect of magnetic field tangling on the observed depolarization. We derive the polarization angle dispersion function denoted by $S$ and calculate the product $P \times S$. The value of $S$ provides an insight into the local non-uniformity in the distributions of the magnetic field morphology and $P \times S$ provides information on the averaged grain alignment efficiency along the line of sight \citep{2020A&A...641A..12P}. A higher value of $S$ implies stronger magnetic field tangling which can decrease the polarization fraction and lower $S$ value implies weaker magnetic field tangling which can result in higher polarization fraction considering a constant grain alignment efficiency along the line-of-sight.

For calculating $S$, we refer to the definition described in Section 3.3 of \cite{2020A&A...641A..12P} and the relation is given below

\begin{equation}
{
S^2(r,\delta) = \frac{1}{N}\sum\limits_{i=1}^{N} {\left[\psi(r+\delta_i) - \psi(r)\right]}^2,
}
\end{equation}
 where the sum extends over the $N$ pixels, indexed by $i$ and located at positions $r+\delta_i$, within a circle centered on $r$ and having radius of $\delta$ taken as two times the beam size of JCMT/POL-2. The term $[\psi(r+\delta_i) - \psi(r)]$ is the difference in the polarization angles at positions $r+\delta_i$ and $r$.

\begin{figure*}
    \centering
    \begin{tabular}{cccc}
        \includegraphics[scale=0.45]{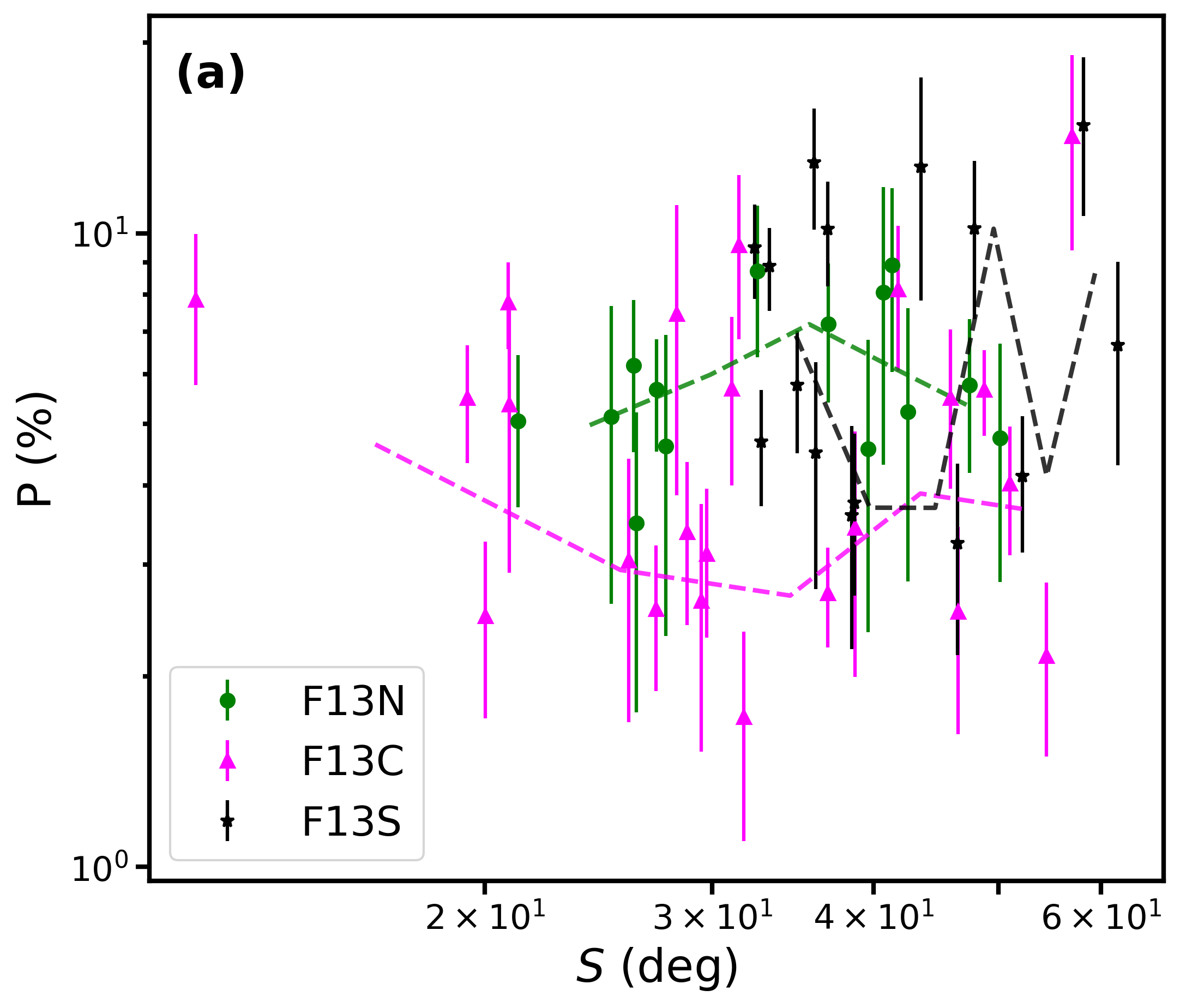} & 
        \hspace{5pt}
        \includegraphics[scale=0.45]{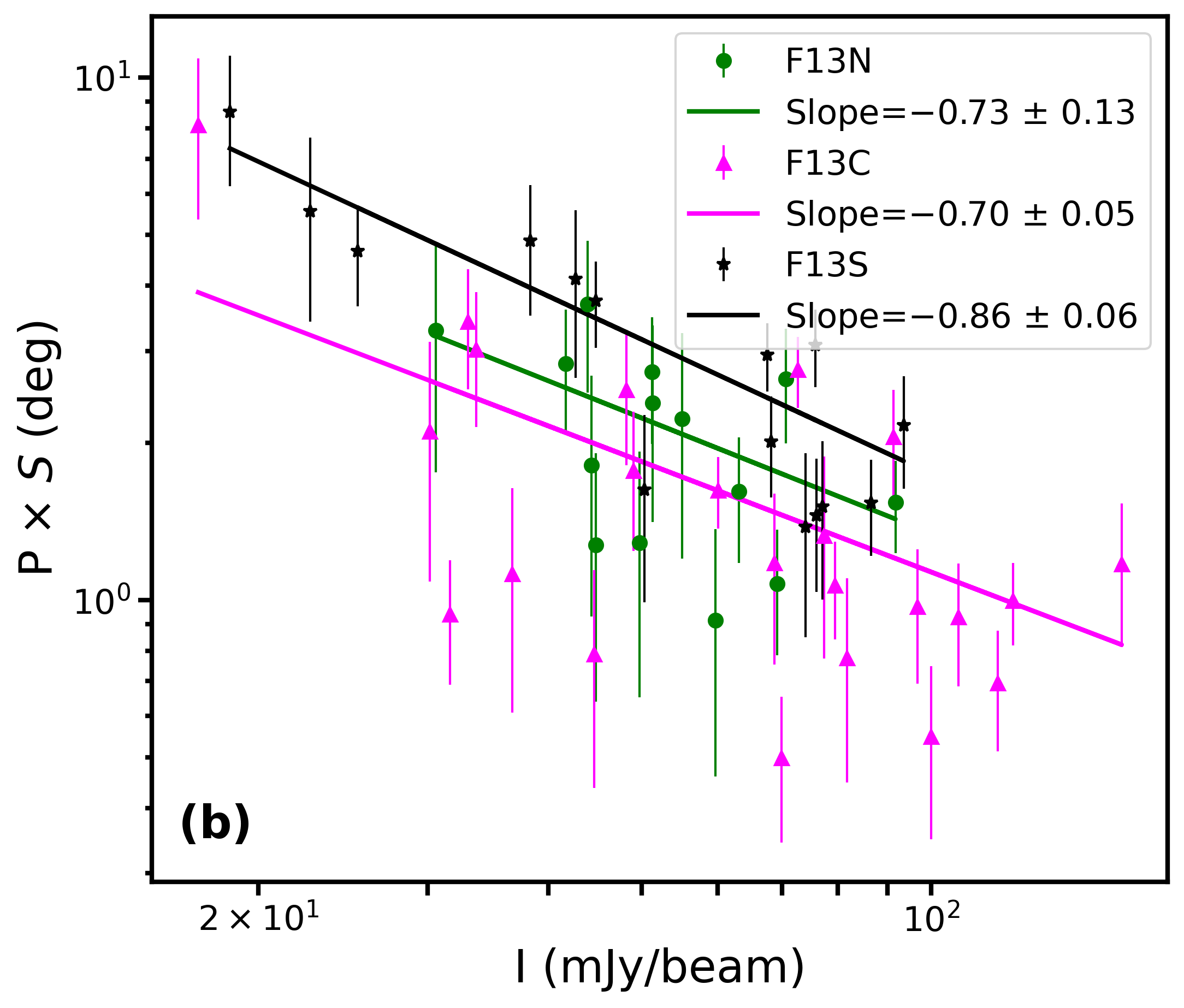} &  \\
        %\hspace{5pt}
        \includegraphics[scale=0.45]{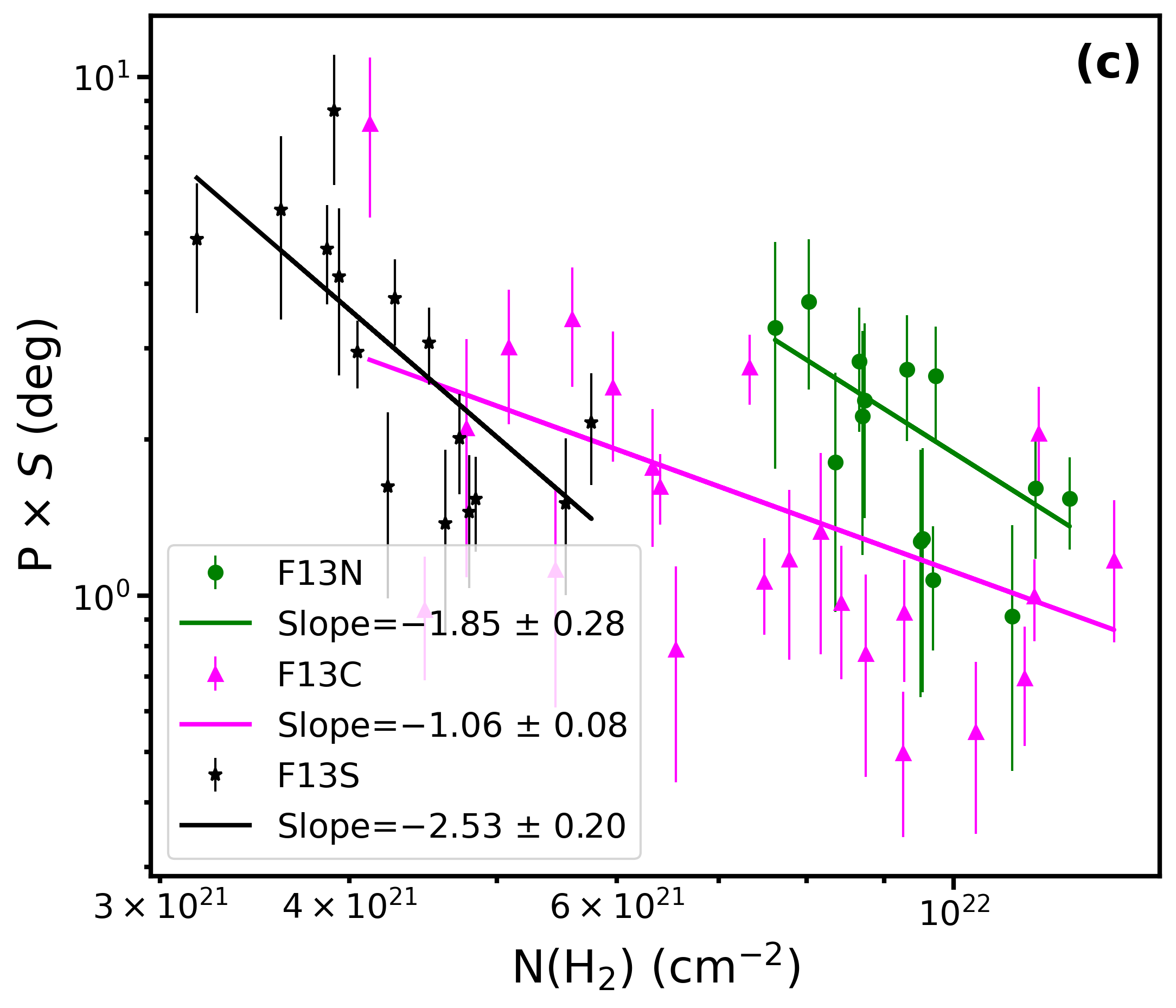} & 
        \hspace{7pt}
        \includegraphics[scale=0.45]{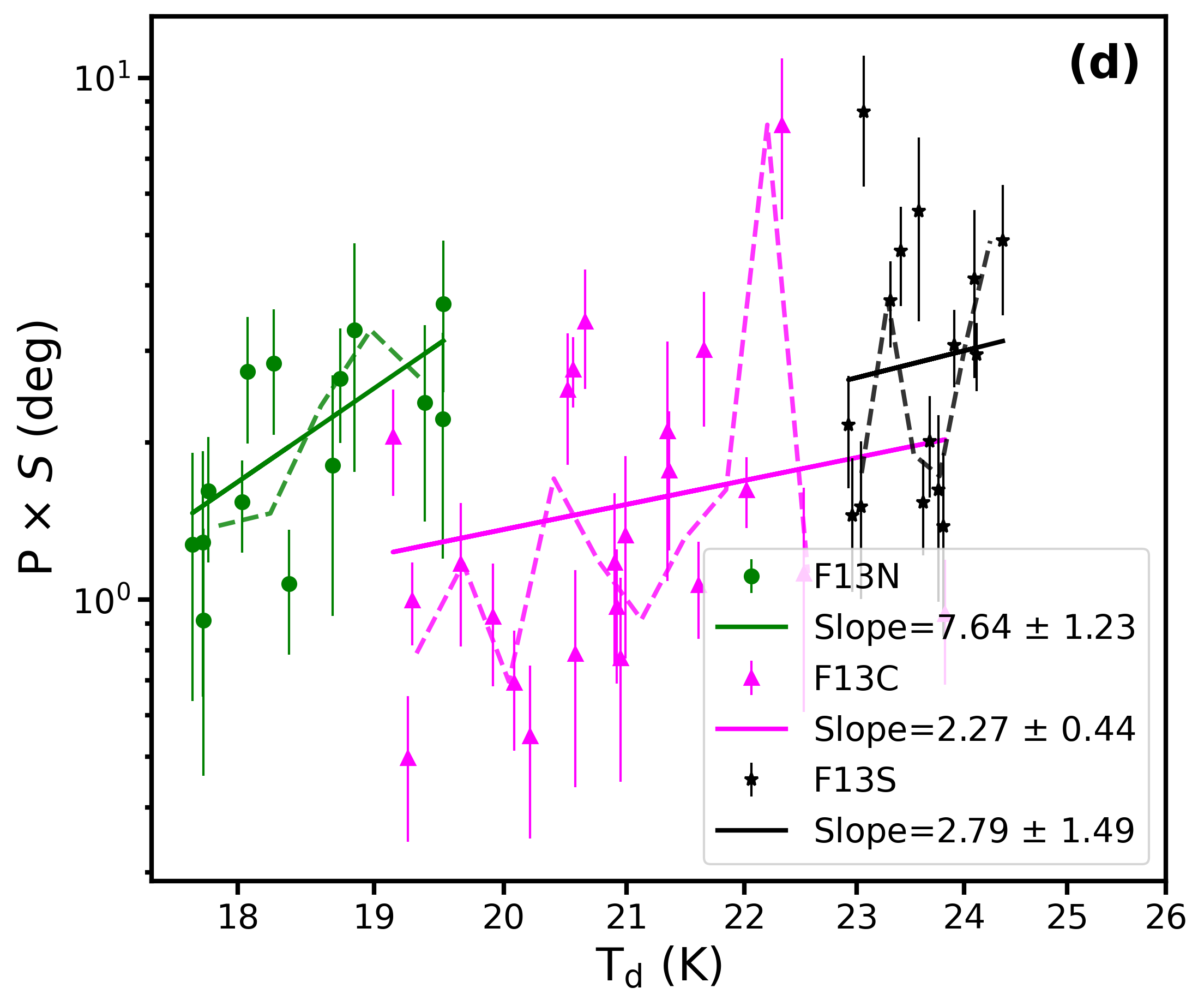}     
    \end{tabular}
    \caption{Variations of (a) $P$ with $S$, (b) $P \times S$ with $I$, (c) $P \times S$ with $N(\mathrm{H_2})$ and (d) $P \times S$ with $T_\mathrm{d}$. The dashed lines are the weighted running means and the solid lines are the weighted best power-law fits.}
    \label{Figure:P-S}
\end{figure*}

As the Stokes parameters $Q$ and $U$ are associated with noise, $S$ becomes biased. This bias of $S$ can be positive or negative depending on whether the true value is smaller or larger than the random polarization angle of $52^\circ$ \citep{2016A&A...595A..57A}. An estimation of the variance of $S$ ($\sigma_S$) resulting from noise along with the debiased values of $S$ ($S_\mathrm{db}$) is described in Section 3.5 of \cite{2020A&A...641A..12P} and are given by the following relations 

\begin{equation}
\begin{split}
% {
&\sigma_S^2(r,\delta)=\frac{\sigma_{\psi}^2(r)}{N^2S^2}\left[\sum\limits_{i=1}^{N} {\psi(r+\delta_i) - \psi(r)}\right]^2 + \\ &\hspace{1.5cm} \frac{1}{N^2S^2}\sum\limits_{i=1}^{N} \sigma_{\mathrm{\psi}}^2(r+\delta_i)\left[\psi(r+\delta_i) - \psi(r)\right]^2
% }
\end{split}
\end{equation}

%\vspace{0.5cm}

and
\begin{equation}
{
S_\mathrm{db}^2(r,\delta) = S^2 - \sigma_S^2 \hspace{0.5cm} \rm{if} \hspace{0.5cm} \it{S > \sigma_S}
}
\end{equation}
We consider only those values of $S_\mathrm{db}$ with $S > \sigma_S$, discarding the other values that do not meet this criterion. Hereafter, we will refer to $S$ as $S_\mathrm{db}(r,\delta)$ for convenience. Then, we examine how $P$ varies with $S$ as shown in Figure \ref{Figure:P-S} (a). Instead of using best-fit lines, we apply weighted running means for each region since the data points exhibit significant spreadness. 

% \begin{figure*}
%     \centering
%     \begin{tabular}{cccc}
%         \includegraphics[scale=0.5]{Cocoon_P_I_Final_2.png} & 
%         \hspace{5pt}
%         \includegraphics[scale=0.5]{Cocoon_P_I_Final_3.png} &  \\
%         %\hspace{5pt}
%         \includegraphics[scale=0.5]{Cocoon_P_NH2_Final_2.png} & 
%         \hspace{5pt}
%         \includegraphics[scale=0.5]{Cocoon_P_NH2_Final_3.png}     
%     \end{tabular}
%     \caption{Variations of (a) $P$ with $I$ for $P/\sigma_P > 2$, (b) $P$ with $I$ for $P/\sigma_P \geq 3$, (c) $P$ with $N(\mathrm{H_2})$ for $P/\sigma_P > 2$, (d) $P$ with $N(\mathrm{H_2})$ for $P/\sigma_P \geq 3$. The white facecolor data points are the data points associated with $2 < P/\sigma_P < 3$. The solid lines are the weighted best power-law fits.}
%     \label{Figure:P_I_NH2}
% \end{figure*}

% \begin{figure*}
%     \centering
%     \begin{tabular}{cc}
%         \includegraphics[scale=0.5]{Cocoon_P_Td_Final_2.png} & 
%         \hspace{5pt}
%         \includegraphics[scale=0.5]{Cocoon_P_Td_Final_3.png} 
%     \end{tabular}
%     \caption{Variations of $P$ with $T_\mathrm{d}$ (a) for $P/\sigma_P > 2$ and (b) for $P/\sigma_P \geq 3$. The white facecolor data points are the data points associated with $2 < P/\sigma_P < 3$. The dashed lines are the weighted running means and the solid lines are the weighted best power-law fits.}
%     \label{Figure:P_Td}
% \end{figure*}

\begin{figure*}
    \centering
    \begin{tabular}{cc}
        \includegraphics[scale=0.4]{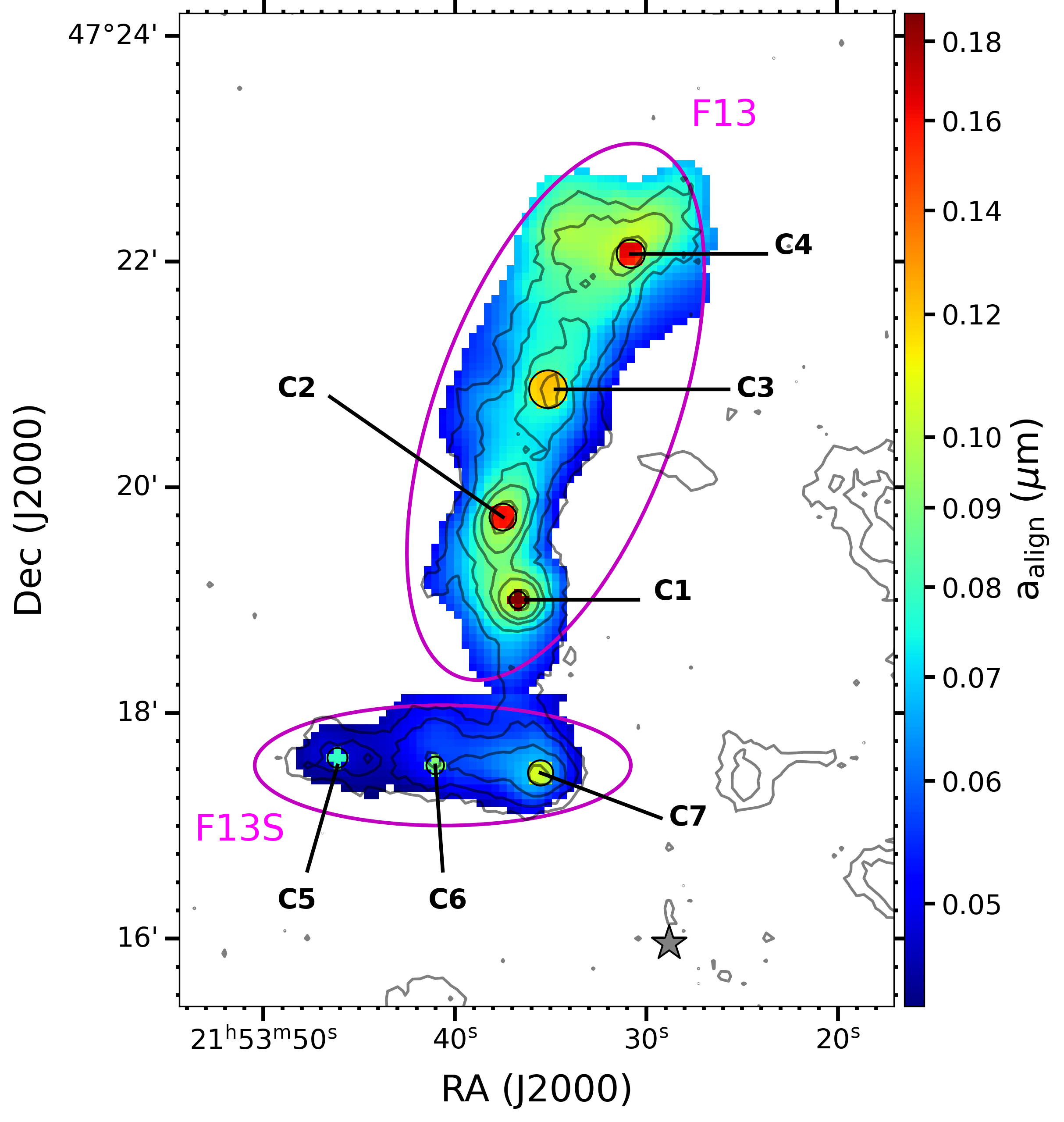} & 
        \hspace{5pt}
        \includegraphics[scale=0.48]{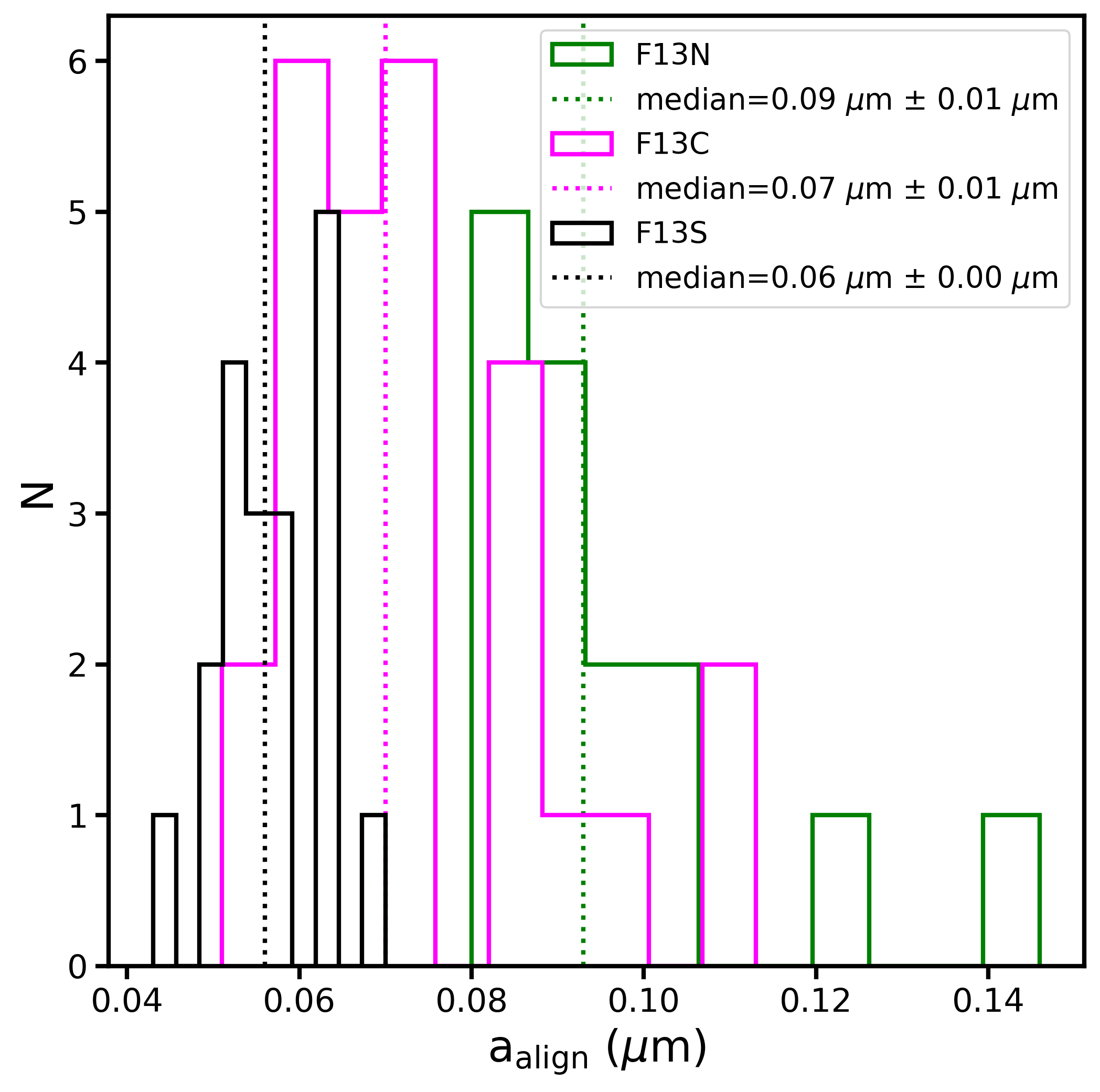} 
    \end{tabular}
    \caption{Map of the minimum alignment size of grains, $a_\mathrm{align}$ (left) and the histograms of $a_\mathrm{align}$ in each of the F13N, F13C and F13S regions (right). The vertical dotted green, magenta and black lines denote the median values of $a_\mathrm{align}$ in F13N, F13C and F13S regions, respectively.}
    \label{Figure:a_align_map}
\end{figure*}

Best-fit lines do not adequately reflect the spreadness in the data, while the weighted running means provide a clearer representation of the variations. We see that there is no much significant correlation between $P$ and $S$ in each of the regions.

Then, we study the variations of the averaged alignment efficiency $P \times S$ with increasing $I$ and $N(\mathrm{H_2})$ as shown in panels (b) and (c) of Figure \ref{Figure:P-S}. We find that $P \times S$ decreases with the increase in both $I$ and $N(\mathrm{H_2})$ nearly similar to the decrease in $P$ with $I$ and $N(\mathrm{H_2})$ which implies that the grain alignment efficiency decreases in the denser regions. Again, we study the variation of $P \times S$ with increasing $T_\mathrm{d}$ as shown in Figure \ref{Figure:P-S} (d) and find that $P \times S$ increases overall in a way more similar to the variation of $P$ with $T_\mathrm{d}$ shown in Figure \ref{Figure:P_Td}. The role of magnetic field tangling to cause the depolarization is found to be small and not much significant. Also, \cite{2024ApJ...970..122C} shows that both the filaments are sub-Alfvenic indicating dominance of magnetic fields over turbulence.

\subsection{Grain alignment mechanism}\label{section:Grain alignment mechanism}
\subsubsection{Minimum alignment size of grains}
The study of grain sizes is of great importance in the context of RAT-A theory to explain the grain alignment mechanism. According to RAT-A theory, efficient alignment of the grains can be achieved only when they rotate suprathermally with a rate exceeding about 3 times the thermal angular velocity (\citealt{2008MNRAS.388..117H, 2016ApJ...831..159H}). At this condition, the randomization of grains by gas-grain collisions can be ignored. The size distribution of aligned grains, spanning from the minimum alignment size, $a_{\mathrm{align}}$ to the maximum size, $a_{\mathrm{max}}$ (\citealt{2014MNRAS.438..680H}; \citealt{2020ApJ...896...44L}) determines the polarization fraction. In order to get an insight into the variations in the polarization fractions in different regions of the filaments in the context of grain sizes, we estimate the values of $a_\mathrm{align}$ over all the filaments using the analytical relation given below as established in \cite{2021ApJ...908..218H}

\begin{equation}
\begin{split}
a_{\mathrm{align}}\simeq0.055\hat{\rho}^{-1/7}\left(\frac{\gamma U}{0.1}\right)^{-2/7}\left(\frac{n_\mathrm{H}}{10^3 \: \mathrm{cm^{-3}}}\right)^{2/7} \\ \times \left(\frac{T_\mathrm{gas}}{10 \: \mathrm{K}}\right)^{2/7} \left(\frac{\bar{\lambda}}{1.2 \: \mu \text{m}}\right)^{4/7} \left(1 + F_\mathrm{IR}\right)^{2/7}, 
\end{split}
\label{equation:a_align} 
\end{equation}
%\vspace{0.3cm}
where $\hat{\rho} = \rho_\mathrm{d}/(3$ $\mathrm{gcm^{-3}})$ with $\rho_\mathrm{d}$ being the dust mass density; $\gamma$ is the anisotropy degree of the radiation field; $\bar{\lambda}$ represents the mean wavelength of the radiation; $U$ is the radiation field strength; $n_\mathrm{H}$ is the number density of hydrogen atoms; $T_\mathrm{gas}$ is the gas temperature and $F_\mathrm{IR}$ is the ratio of the IR damping to the collisional damping rate. For the diffuse ISRF, $\gamma = 0.1$ (\citealt{1997ApJ...480..633D}; \citealt{2007ApJ...663.1055B}). However, in our study we take $\gamma = 0.3$ as a B-type star BD+46 is present nearby and hence the radiation becomes more anisotropic in the elongated dense filamentary clouds. We use $\rho_\mathrm{d}=3$ $\mathrm{gcm^{-3}}$, $\bar{\lambda}=1.2$ $\mu$m, $n_\mathrm{H} = 2n(\mathrm{H_2})$ where $n(\mathrm{H_2})$ is the volume density of molecular hydrogen gas and $T_\mathrm{gas}=T_\mathrm{d}$ is considered as this thermal equilibrium between gas and dust is valid for dense and cold environments. For dense molecular clouds, $F_\mathrm{IR} << 1$. 

\begin{figure}
    \centering
        \includegraphics[scale=0.45]{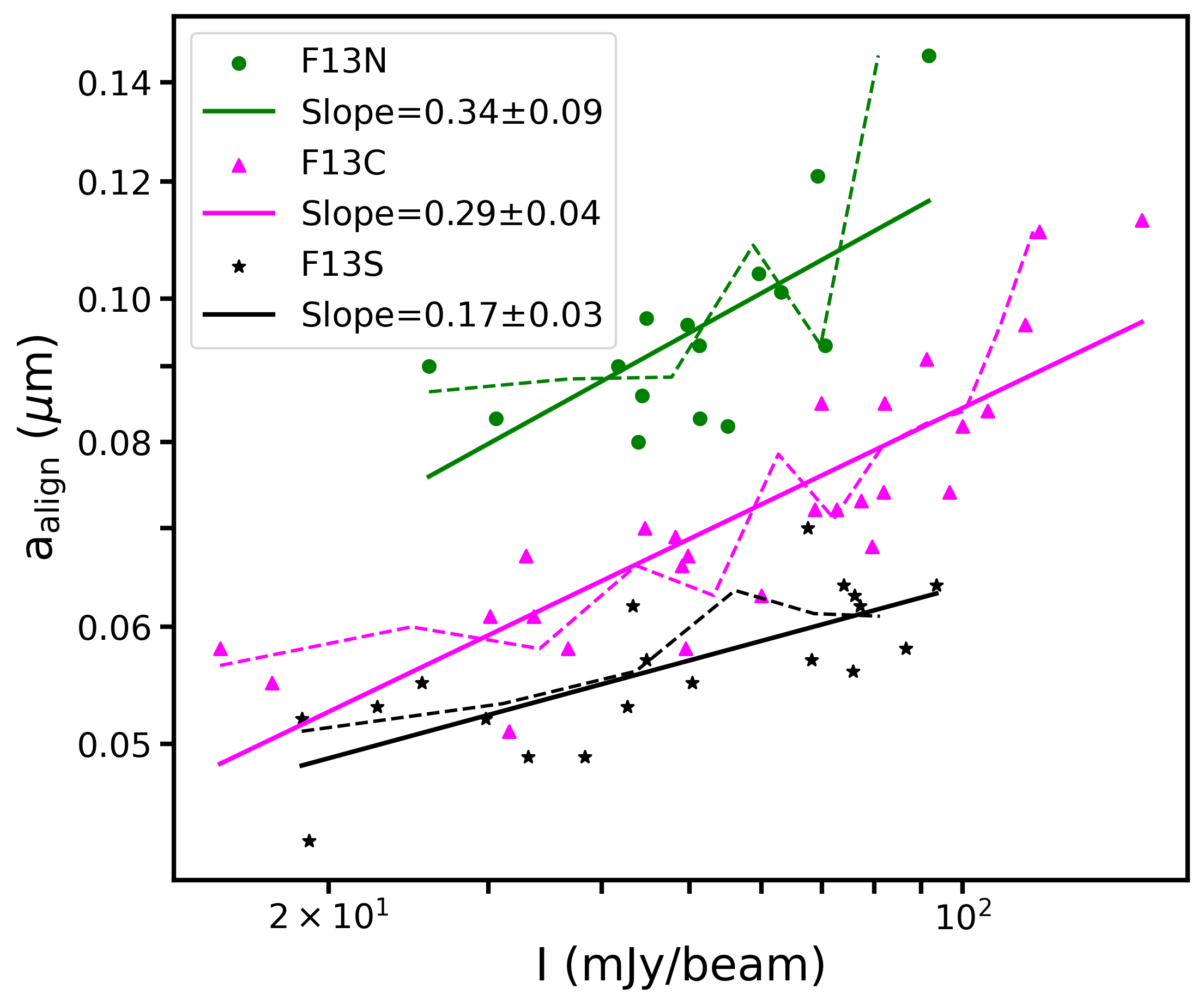} 
    \caption{Variations of $a_\mathrm{align}$ with the total intensity $I$ in each of the regions. The dashed lines are the running means and the solid lines are the best power-law fits.}
    \label{Figure:a_align_I}
\end{figure}

\begin{figure*}
    \centering
    \begin{tabular}{cc}
        \includegraphics[scale=0.45]{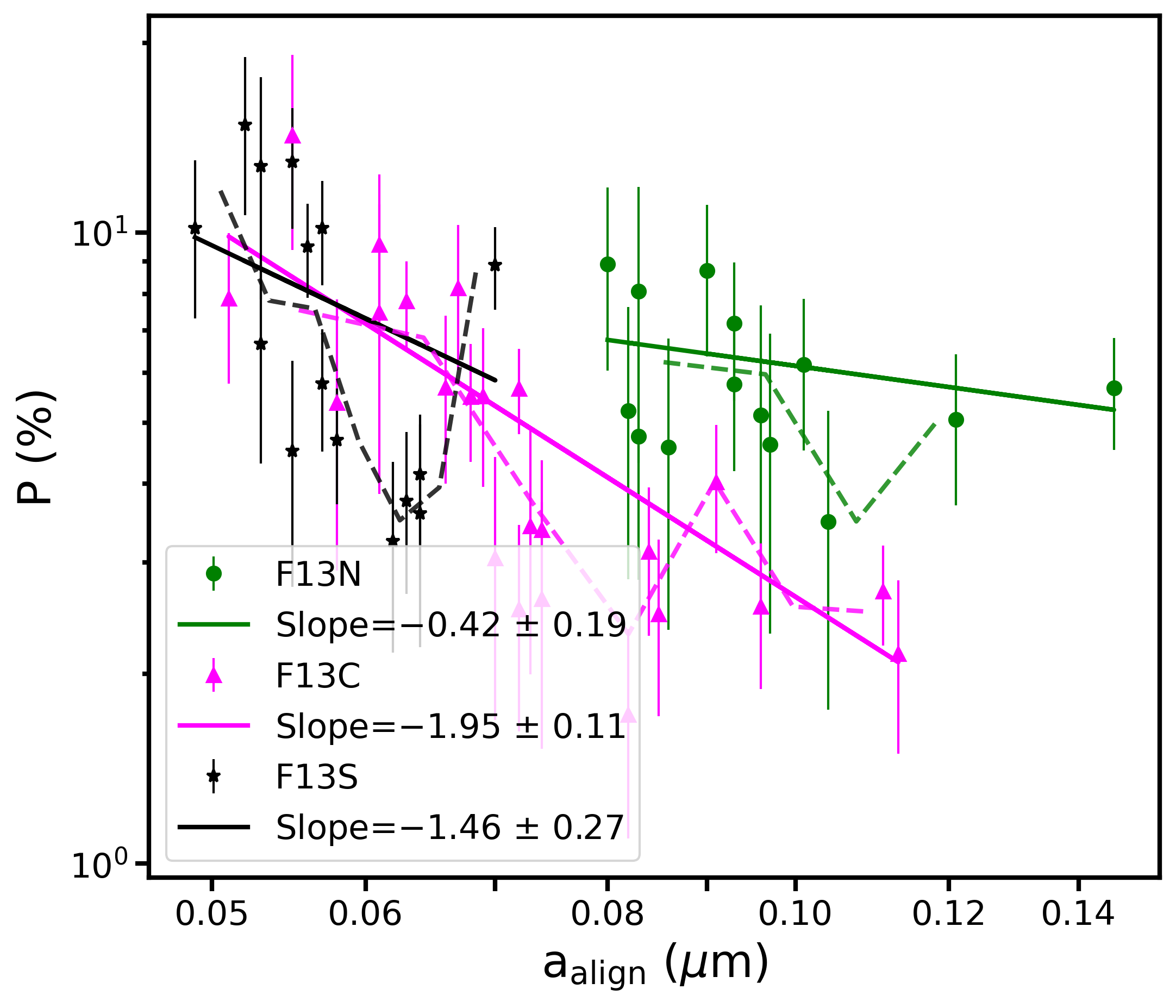} & 
        \hspace{5pt}
        \includegraphics[scale=0.45]{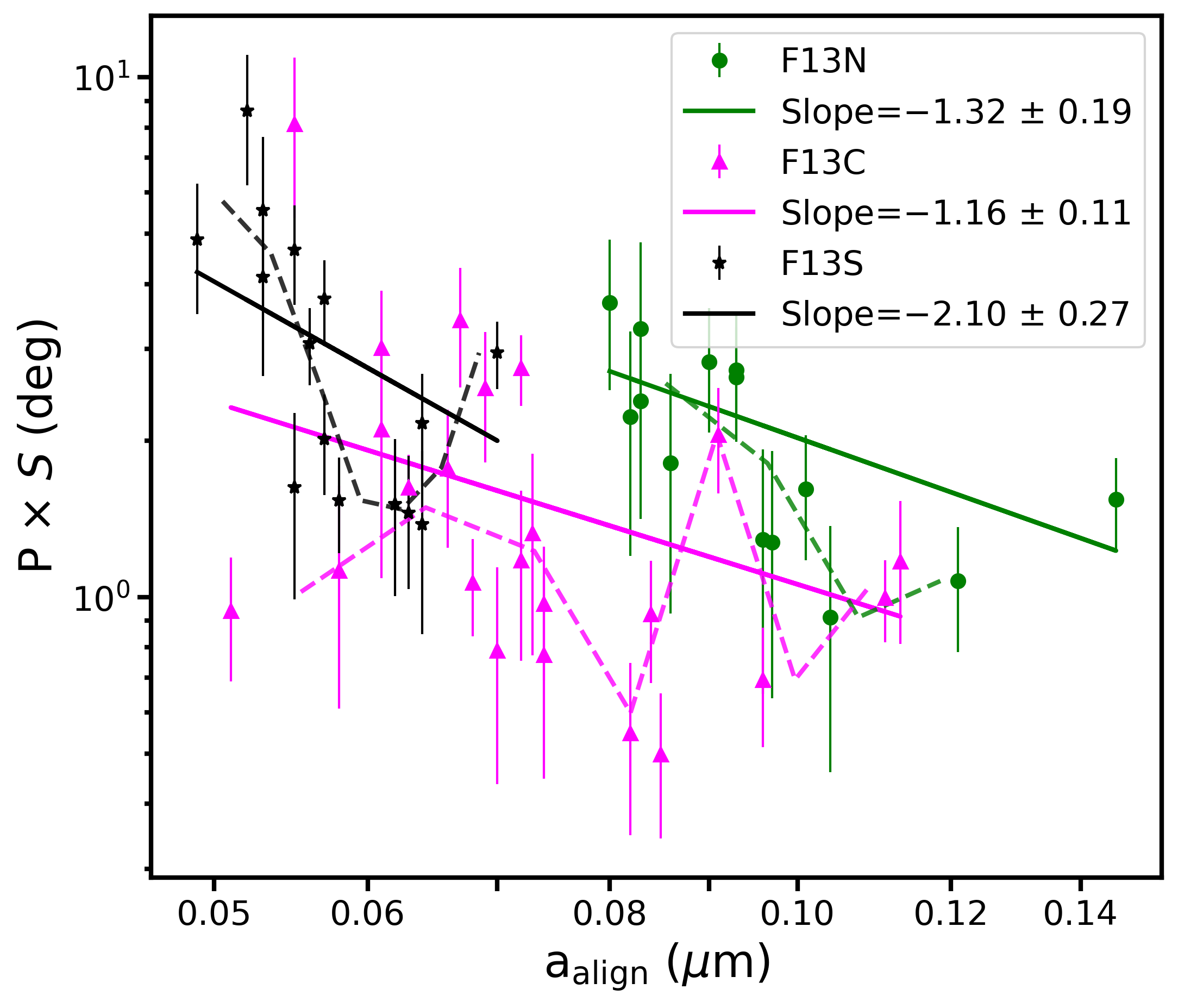} 
    \end{tabular}
    \caption{Variations of $P$ with $a_\mathrm{align}$ (left) and $P \times S$ with $a_\mathrm{align}$ (right). The dashed lines are the weighted running means and the solid lines are the weighted best power-law fits.}
    \label{Figure:P_P_S_a_align}
\end{figure*}

\setlength{\tabcolsep}{1.2cm} 
\renewcommand{\arraystretch}{1.6}

\begin{table*}[ht]
    \centering
        \caption{Slope values of weighted best power-law fits for different analyses using $P/\sigma_{P} > 2$}
    \begin{tabular}{|c|c|c|c|}
        \hline
        
        Relation between & F13N & F13C & F13S\\ \hline
        $P$ vs. $I$ & $-0.71 \pm 0.09$ & $-0.94 \pm 0.04$ & $-0.68 \pm 0.05$ \\ \hline
        $P$ vs. $N(\mathrm{H_2})$ & $-1.61 \pm 0.23$ & $-1.70 \pm 0.07$ & $-1.73 \pm 0.14$ \\ \hline
        $P \times S$ vs. $I$ & $-0.73 \pm 0.13$  & $-0.70 \pm 0.05$ & $-0.86 \pm 0.06$  \\ \hline
        $P \times S$ vs. $N(\mathrm{H_2})$ & $-1.85 \pm 0.28$  & $-1.06 \pm 0.08$ & $-2.53 \pm 0.20$  \\ \hline
        $P$ vs. $T_\mathrm{d}$ & $-1.37 \pm 1.16$ & $7.89 \pm 0.37$ & $5.57 \pm 0.86$ \\ \hline
        $P \times S$ vs. $T_\mathrm{d}$  & $7.64 \pm 1.23$ & $2.27 \pm 0.44$ & $2.79 \pm 1.49$  \\ \hline
        $a_\mathrm{align}$ vs. $I$ & $0.34 \pm 0.09$ & $0.29 \pm 0.04$ & $0.17 \pm 0.03$   \\ \hline
        $P$ vs. $a_\mathrm{align}$ & $-0.42 \pm 0.19$ & $-1.95 \pm 0.11$ & $-1.46 \pm 0.27$   \\ \hline
        $P \times S$ vs. $a_\mathrm{align}$ & $-1.32 \pm 0.19$ & $-1.16 \pm 0.11$ & $-2.10 \pm 0.27$ \\ \hline          
    \end{tabular}
    \label{Table:Table_1}
\end{table*}

To calculate $U$, we use the relation between dust temperature and the radiation strength for silicate grains having sizes in the range of 0.01-1$\mu$m with dust heating and cooling balance and radiation strength $U < 10^4$ $(\approx 75$ K) i.e $U$ $\approx$ $(T_\mathrm{d}/16.4$ $\mathrm{K})^6$ \citep{2011piim.book.....D}. The map of the alignment size and the histogram distributions are shown in the left and the right panels of Figure \ref{Figure:a_align_map}. The median values of $a_\mathrm{align}$ for the F13N, F13C and F13S regions are found to be $0.09 \pm 0.01$ $\mu$m, $0.07 \pm 0.01$ $\mu$m and 0.06 $\mu$m and are indicated in the figure with green, magenta and black vertical dotted lines respectively. The histogram shows that the F13N and F13C regions have distributions of larger $a_\mathrm{align}$ values compared to F13S region which shows distribution of smaller $a_\mathrm{align}$ values. Equation \ref{equation:a_align} shows that $a_\mathrm{align}$ varies with the radiation field strength $U$ or equivalently the dust temperature $T_\mathrm{d}$ as $U^{-2/7}$ and with the gas volume density $n_\mathrm{H}$ as $n_\mathrm{H}^{2/7}$. From the map of $a_\mathrm{align}$, we find that the F13N region shows a high value of $a_\mathrm{align}$. This F13N region has high volume density and low dust temperature (see Figure \ref{Figure:Volume_density_Td_map}). The densest core regions of C1 and C2 in F13C region show very high $a_\mathrm{align}$ values compared to other core regions. The F13S region is relatively less denser and has higher dust temperature compared to other regions and the $a_\mathrm{align}$ values are less. However, overall $a_\mathrm{align}$ increases from the outer regions towards the inner regions in both the filaments.

Within the framework of the RAT-A theory, the size distributions of aligned grains range from $a_\mathrm{align}$ to $a_\mathrm{max}$, and the polarization fraction is determined by this range of grain size distributions. The processes of grain growth and destruction determine the value of $a_\mathrm{max}$. When $a_{\mathrm{max}}$ is fixed, an increase in the value of $a_{\mathrm{align}}$ can provide narrower size distribution of aligned grains which can result in the reduction of the polarization fraction $P$. Again, a decrease in $a_{\mathrm{align}}$ can result in wider size distribution of aligned grains which can increase $P$ (see Figure 7 in \citealt{2022FrASS...9.3927T}). Therefore, an anti-correlation is expected between $a_{\mathrm{align}}$ and $P$. Also, we expect $a_\mathrm{align}$ to increase with the total intensity $I$ in starless clouds.

Figure \ref{Figure:a_align_I} shows the variation of $a_{\mathrm{align}}$ with $I$. We see that $a_\mathrm{align}$ linearly increases with $I$ in each of the regions. Figure \ref{Figure:P_P_S_a_align} shows the variation of $P$ (left) and $P \times S$ (right) with $a_{\mathrm{align}}$. We plot weighted power-law fits and also weighted running means in each region. We see $P$ decreases with the increase in $a_\mathrm{align}$ in each of the regions and $P \times S$ also decreases with $a_\mathrm{align}$ nearly similar to the decrease of $P$ with $a_\mathrm{align}$ which imply that grain alignment efficiency decreases as $a_\mathrm{align}$ increases. Polarization fraction decreases as $a_{\mathrm{align}}$ increases due to the reduction in the fraction of aligned grains, from numerical modeling (\citealt{2020ApJ...896...44L}; \citealt{2021ApJ...908..218H}). The depolarization in each of the regions could be due to the decrease in the RAT alignment efficiency of grains in the denser regions of high column densities and low dust temperatures.

Table \ref{Table:Table_1} summarizes the slope values of the weighted best power-law fits for each of the above different analyses.

\subsection{Effect of magnetic relaxation on the RAT Alignment}

In the investigation of grain alignment, the dust magnetic properties are very crucial as these properties enable the grains to interact with the external magnetic field. When there is diffuse distribution of iron atoms within a silicate grain, the grain behaves as an ordinary paramagnetic material. However, when iron atoms are distributed as clusters, the grain becomes super-paramagnetic \citep{2022AJ....164..248H}. A paramagnetic grain that rotates with an angular velocity $\omega$ in the presence of an external magnetic field $B$ undergoes paramagnetic relaxation \citep{1951ApJ...114..206D} that induces dissipation of the rotational energy of the grains into heat, resulting in the gradual alignment of angular velocity and angular momentum with $B$, known as the classical Davis-Greenstein mechanism which is applicable to any magnetic material. However, the efficient alignment of grains can not be achieved by the paramagnetic relaxation alone due to randomization of grains by gas-grain collisions. Again, only RATs cannot produce perfect alignment of grains as the RAT alignment efficiency depends on various factors such as the angle between the radiation field direction and the magnetic field direction, grain properties like shape and compositions (\citealt{2016ApJ...831..159H}; \citealt{2021ApJ...913...63H}). In the filaments of our study, it is observed that there are $P$ values of more than 5\% and reaching up to around 15\%. Therefore, we explore the effect of the magnetic relaxation on the RAT Alignment efficiency by considering super-paramagnetic grains having embedded iron atoms as clusters, to explain the observed high $P$ values, especially in the outer regions of the filaments. This consideration of super-paramagnetic nature of the grains is expected in denser regions due to the evolution of grains.

A dimensionless parameter $\delta_\mathrm{mag}$ was introduced in \cite{2016ApJ...831..159H} to describe the aligning effect of magnetic relaxation relative to the disalignment caused by gas collisions. This parameter $\delta_\mathrm{mag}$ provides the strength of the magnetic relaxation and is defined as the ratio of the gas collision damping timescale, $\tau_\mathrm{gas}$ to the magnetic relaxation time, $\tau_\mathrm{mag,sp}$. For super-paramagnetic grains having embedded iron atoms as clusters, the strength of the magnetic relaxation is given by the following relation
\vspace{0.2cm}
\begin{equation}
{
\delta_\mathrm{mag,sp} = \frac{\tau_\mathrm{gas}}{\tau_\mathrm{mag,sp}} = 56a^{-1}_{-5} \frac{N_\mathrm{cl} \phi_\mathrm{sp,-2} \hat{p}^2 B_3^2}{\hat{\rho} n_4 T_\mathrm{gas,1}^{1/2}} \frac{k_\mathrm{sp}(\Omega)}{T_\mathrm{d,1}}, \label{equation:magnetic relaxation}
}
\end{equation} 
%\vspace{0.3cm}
where $a_{-5}=a/(10^{-5}$ cm), $B_3=B_\mathrm{tot}/(10^3$ $\mu$G), $n_4=n_\mathrm{H}/(10^4$ $\mathrm{cm^{-3}}$) with $n_\mathrm{H} \approx 2n(\mathrm{H_2})$ for molecular gas, $T_\mathrm{gas,1}=T_\mathrm{gas}/(10$ K), $T_\mathrm{d,1}=T_\mathrm{d}/(10$ K), $\hat{p}=p/5.5$ with $p \approx 5.5$ the coefficient describing the magnetic moment of an iron atom, $N_\mathrm{cl}$ gives the number of iron atoms per cluster, $\phi_\mathrm{sp}$ is the volume filling factor of iron clusters with $\phi_\mathrm{sp,-2}=\phi_\mathrm{sp}/0.01$ and $k_\mathrm{sp}(\Omega)$ is the function of the grain rotation frequency $\Omega$ which is of order unity \citep{2022AJ....164..248H}.

When the magnetic relaxation occurs much faster than the gas collision damping, the magnetic relaxation strength is considered to be effective for the alignment of grains. The degree of RAT alignment of grains can be significantly enhanced by the combined effect of both the suprathermal rotation of grains by RATs and the strong magnetic relaxation strength, which is termed as the Magnetically-enhanced RAdiative Torque (M-RAT) mechanism of grain alignment.

\begin{figure*}
    \centering
    \begin{tabular}{cc}
        \hspace{-35pt}
        \includegraphics[scale=0.35]{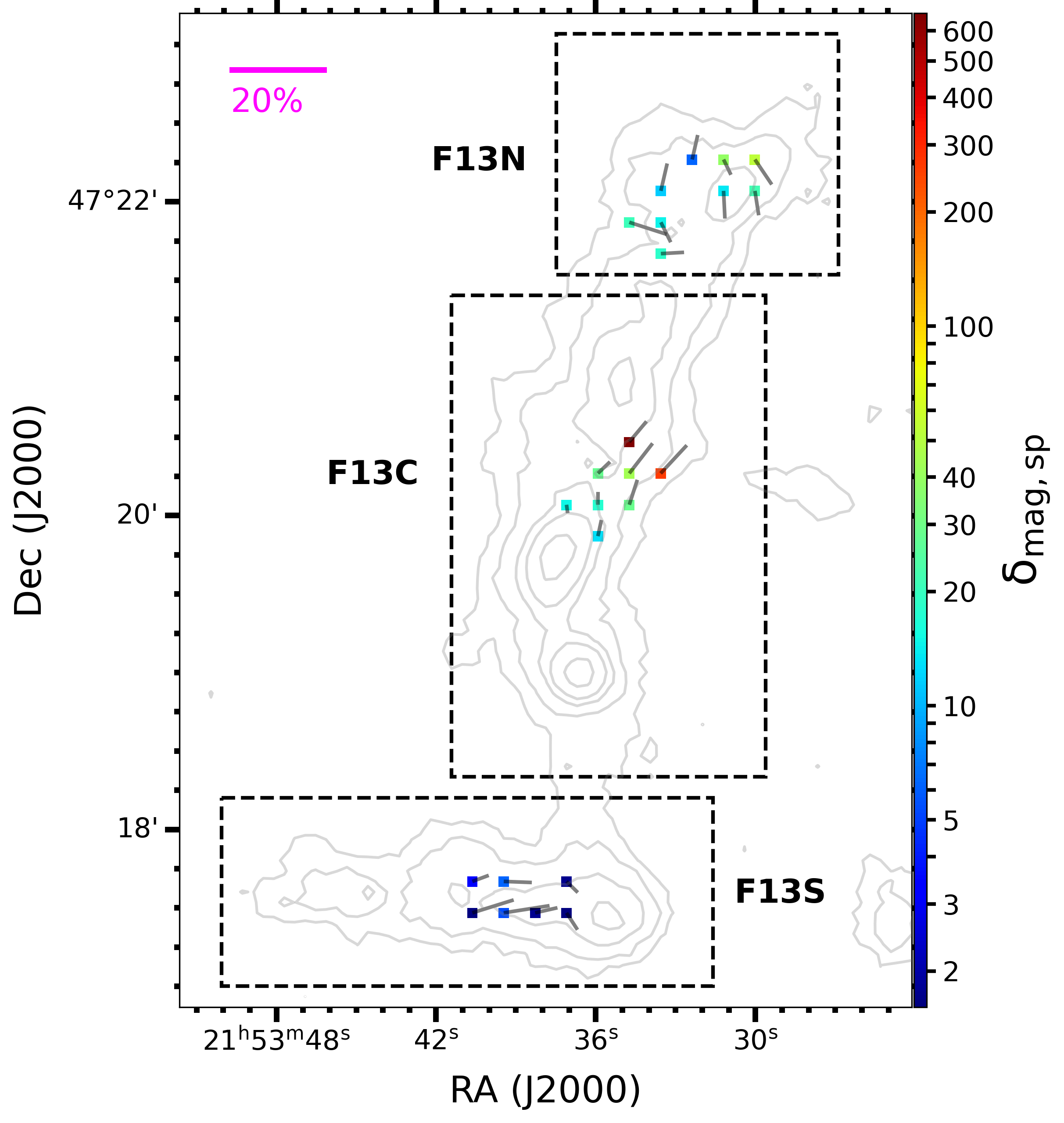} & 
        \hspace{-65pt}
        \includegraphics[scale=0.38]{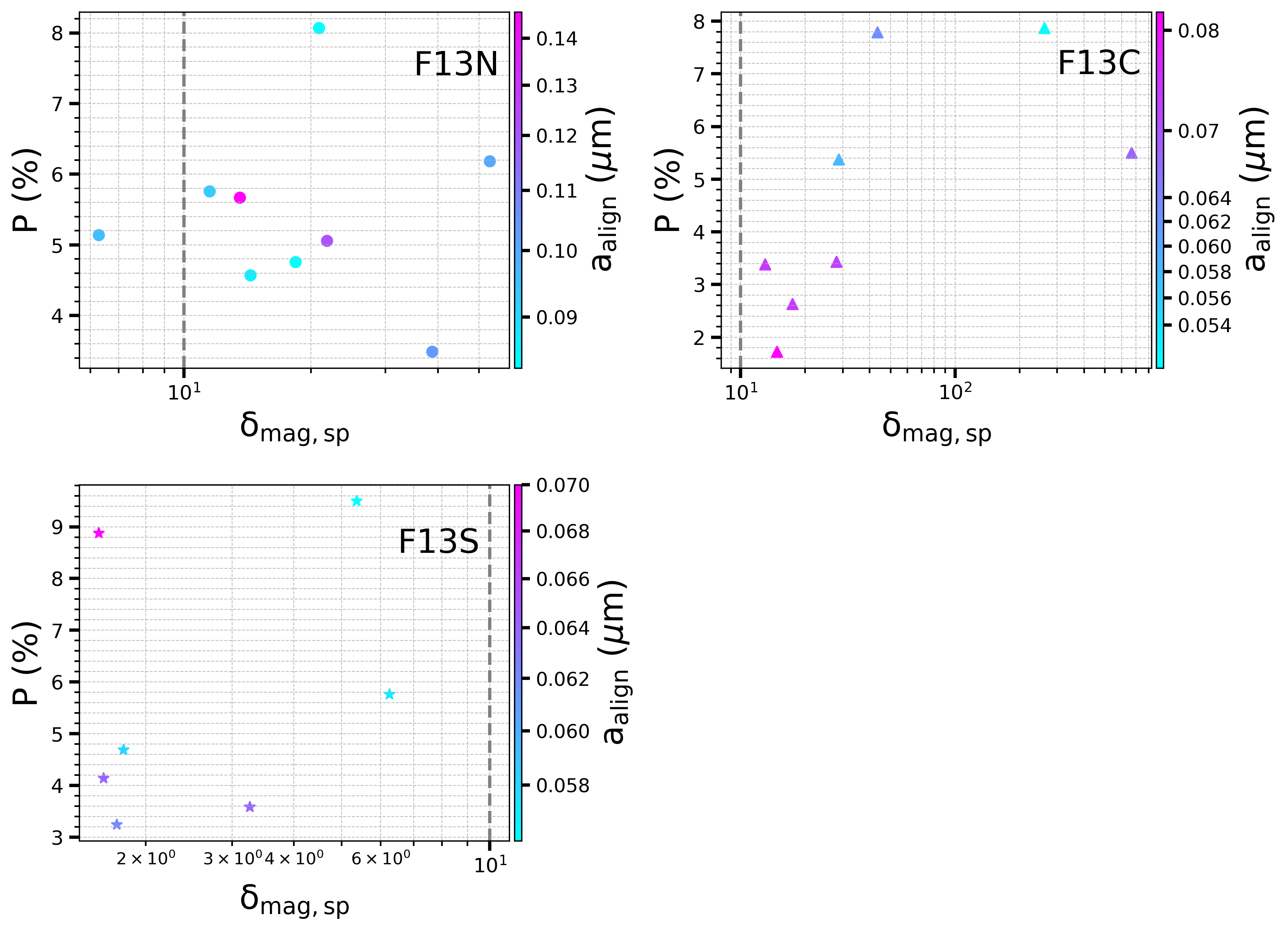} 
    \end{tabular}
    \caption{Map of magnetic relaxation strength $\delta_\mathrm{mag,sp}$ with overlaid gray polarization vectors (left) and the relation between polarization fraction $P$ and $\delta_\mathrm{mag,sp}$ (right) in each region for only 24 pixels in total having estimated values of magnetic field strengths. The length of the gray vectors in the left panel are proportional to the polarization fraction $P$. A reference scale of $P$ is also indicated. The vertical dashed gray lines denote $\delta_\mathrm{mag,sp} = 10$. Most of the pixels in the F13N and F13C regions show $\delta_\mathrm{mag,sp} > 10$. Some pixels have $\delta_\mathrm{mag,sp} > 10$ with corresponding higher $P$ values of around $5-8$\% in the F13N and F13C regions. The pixels in F13S region have $\delta_\mathrm{mag,sp} << 10$.}
    \label{Figure:mr_map}
\end{figure*}

To calculate the magnetic relaxation strength $\delta_\mathrm{mag,sp}$ values in all regions of the F13 and F13S filaments using Equation \ref{equation:magnetic relaxation}, we use the pixel values of the plane-of-sky magnetic field strength $B_\mathrm{POS}$ from the $B_\mathrm{POS}$ pixel-by-pixel map as given in Figure 16 (left) in \cite{2024ApJ...970..122C}. For the details on the estimation of the $B_\mathrm{POS}$ map, please refer to Figures 12, 15 and 16 in \cite{2024ApJ...970..122C}. In this map, $B_\mathrm{POS}$ values are estimated only for 24 detected pixels, which lacks $B_\mathrm{POS}$ information for the other pixels having polarization data. Therefore, our calculation of $\delta_\mathrm{mag,sp}$ will be limited to only these pixels. We calculate the total magnetic field strength $B_\mathrm{tot}$ by multiplying $B_\mathrm{POS}$ by a factor of 1.3 \citep{2004ApJ...600..279C}. We use $N_\mathrm{cl}=100$ and $\phi_\mathrm{sp}=0.01$ (about 3\% of iron abundance as iron clusters, \citealt{2016ApJ...831..159H}). We calculate the $\delta_\mathrm{mag,sp}$ values for those 24 pixels using the values of all the parameters in Equation \ref{equation:magnetic relaxation}. The map of the strength of magnetic relaxation is shown in the left panel of Figure \ref{Figure:mr_map}. The relation between the polarization fraction $P$ and $\delta_\mathrm{mag,sp}$ for each region with the colors denoting the $a_\mathrm{align}$ values is shown in the right panel of Figure \ref{Figure:mr_map}. In the F13N region, the values of $\delta_\mathrm{mag,sp}$ are in the range from $6-53$ with a median value of around 18, in the F13C region the values are in the range from $13-667$ with a median value of around 28 and in the F13S region from $2-6$ with a median value of around 2. 
We find some pixels associated with high $B_\mathrm{POS}$ values in the F13C region and having strong magnetic relaxation strength $\delta_\mathrm{mag,sp} >> 10$. The $P$ values in the corresponding pixels are also found to be high ranging from around 5\% to 8\%. We expect that there may not be strict correlation between $P$ and $\delta_\mathrm{mag,sp}$ as the pixels are associated with different $a_\mathrm{align}$ values and $P$ also depends on $a_\mathrm{align}$ values. In F13C, we see the increase in $P$ as $\delta_\mathrm{mag,sp}$ increases and $a_\mathrm{align}$ decreases in Figure \ref{Figure:mr_map}. The F13N region also has $\delta_\mathrm{mag,sp} > 10$ in some pixels with high $P$ values ranging from 5\% to 8\%. However, we see some fluctuations in $P$ values in F13N region. The F13S region has small $B_\mathrm{POS}$ values overall and $\delta_\mathrm{mag,sp} << 10$ (for a discussion see Section \ref{section:Discussion_magnetic_relaxation}). More observational data are needed to better understand the role of enhanced magnetic relaxation on the RAT alignment or the M-RAT mechanism in these filaments.

\section{Discussions}\label{section:Discussions}
In this section, we will discuss our results to explain the mechanism of grain alignment in each of the regions of the filaments.

\subsection{Evidence for the alignment of grains by RAT-A mechanism}\label{section:RAT-A mechanism}
\subsubsection{Decrease in the polarization fraction with the total intensity and gas column density}\label{section:Discussions_P-I-NH2}
Observational studies of dust polarization in various environments from the diffuse interstellar medium to the molecular clouds, filaments and star-forming regions reveal that the polarization fraction decreases with the increase in the total intensity and gas column density, widely known as polarization hole (see e.g, \citealt{2019FrASS...6...15P}; \citealt{2021ApJ...908..218H}). However, the exact cause of the polarization hole is still not clear. The significant decrease in the grain alignment in the denser regions having no embedded sources, by RATs for uniform magnetic fields act as the leading explanation for the polarization hole \citep{2021ApJ...908..218H}. The polarization hole can also be caused by the fluctuations in the magnetic field orientations along the line of sight. These fluctuations in the magnetic fields are suggested to be due to turbulence in the region (\citealt{1989ApJ...346..728J}; \citealt{1992ApJ...389..602J}; \citealt{2008ApJ...679..537F}).  

The F13 and F13S filaments does not have prominent bright embedded sources inside. However, there is a single B-type star BD+46 in the Southern part of F13 filament. This environmental condition makes this region favourable to study the grain alignment mechanism in the context of RAT theory. The diffuse ISRF and the radiation field from the nearby star BD+46 act as the main radiation source for grain heating and alignment in these regions. 

In the $P-I$ and $P-N(\mathrm{H_2})$ plots (see Figure \ref{Figure:P_I_NH2}), the polarization fraction is found to decrease with the increase in the total intensity and the gas column density. Again, in the $T_\mathrm{d}-N(\mathrm{H_2})$ plot (see Figure \ref{Figure:Td_NH2}), the dust temperature decreases with the increase in the gas column density which implies the absence of bright embedded sources. The only source of grain heating is from the nearby star BD+46 and the diffuse ISRF and the external radiation field strength decreases in the denser regions. Since the F13 and F13S filaments lack bright embedded sources, the high gas density in the denser regions dominate over the weak radiation strength and randomize the grains due to gas-grain collisions resulting in the decrease in grain alignment efficiency in these denser regions. We see high values of polarization fraction in the outer regions of the filaments up to around 15\%. This may be due to the efficient alignment of grains in these outer regions by the strong radiation field from the BD+46 star. However, this radiation will get attenuated significantly as it traverse through the denser regions.

We also show that the magnetic field tangling does not have much significant effect on the observed depolarization from our analyses of the variations of $P$ with $S$, $P$ and $P \times S$ with $I$, $N(\mathrm{H_2})$ and $T_\mathrm{d}$ (see Figure \ref{Figure:P-S}). Hence, the depolarization in the denser regions is majorly due to the decrease in the alignment efficiency of grains in the denser regions in agreement with the RAT theory. The similar observed feature that the depolarization in the denser regions is predominantly due to significant decrease in grain alignment efficiency, with the B-field tangling acting as sub-dominant and less significant role was also found in the study in dense and cold filaments which do not have prominent embedded bright sources, the G11.11$-$0.12 by \cite{2023ApJ...953...66N} and the Musca by \cite{2024ApJ...974..118N}. Hence, it provides us a possible implication that this feature may be universal in dense and cold filaments. However, we need to explore more other dense filamentary regions to strengthen the implication of being universality of this observed feature in dense filaments.   

\subsubsection{Increase in the polarization fraction with dust temperature}\label{section:Discussions_P-Td}
Another prediction of the RAT theory is that the polarization fraction should increase with the increase in the dust temperature (see e.g., \citealt{2020ApJ...896...44L}; \citealt{2021ApJ...908..218H}). In each of the regions of the filaments, both the polarization fraction $P$ and the averaged grain alignment efficiency $P \times S$ increases with the increase in the dust temperature overall (see Figures \ref{Figure:P_Td} and \ref{Figure:P-S} (d)). This observed feature agrees well with the expectation of the RAT theory and provides us another evidence for the RAT alignment mechanism of grains.

\subsubsection{Increase in $a_{align}$ with I; decrease in P and $P \times S$ with $a_{align}$}\label{section:Discussions_P-a_align}
Using the RAT-A theory, we calculate the minimum size for aligned grains by RATs using the local physical parameters and derive a map of alignment size (see Figure \ref{Figure:a_align_map}). Within the framework of the RAT-A theory, the size distributions of aligned grains range from $a_\mathrm{align}$ to $a_\mathrm{max}$, and the polarization fraction is determined by this range of grain size distributions. When $a_{\mathrm{max}}$ is fixed, an increase in the value of $a_{\mathrm{align}}$ can provide narrower size distribution of aligned grains which can result in the reduction of the polarization fraction $P$. Again, a decrease in $a_{\mathrm{align}}$ can result in wider size distribution of aligned grains which can increase $P$ (see Figure 7 in \citealt{2022FrASS...9.3927T}). Therefore, an anti-correlation is expected between $a_{\mathrm{align}}$ and $P$. Also, we expect $a_\mathrm{align}$ to increase with the total intensity $I$ in starless clouds.

We find that the value $a_\mathrm{align}$ is well correlated with the intensity (see Figure \ref{Figure:a_align_I}) which means the alignment size increases in denser regions. The increasing of alignment size with intensity when no internal radiation source is present is an expectation of RAT-A theory. Also, there is anti-correlation in $P$ with $a_{\mathrm{align}}$ and $P \times S$ with $a_{\mathrm{align}}$ (see Figure \ref{Figure:P_P_S_a_align}). These observed features well support the RAT-A theory.

\subsection{Role of magnetic relaxation on the RAT Alignment: M-RAT mechanism}\label{section:Discussion_magnetic_relaxation}
The study of the effect of enhanced magnetic relaxation due to the expected super-paramagnetic grains in the denser regions, on the RAT Alignment of grains is important in the investigation of the exact alignment mechanism of grains. The observational potential evidence for the alignment of grains by the combined effects of both the suprathermal rotation by RATs and the enhanced magnetic relaxation strength or M-RAT mechanism was found in the previous studies in G11.11$-$0.12 filament by \cite{2023ApJ...953...66N} and in G34.43+0.24 filament by \cite{2025arXiv250111634P}. In the F13 and F13S filaments, we find some pixels having high $P$ values ranging from $6-15$\%. This observed high $P$ values may not be produced by RAT only due to its dependence on factors like the angle between the radiation field direction and the magnetic field direction, grain shapes and compositions, etc. (\citealt{2016ApJ...831..159H}; \citealt{2021ApJ...913...63H}). In our study, due to the limitation on the information of the estimated $B_\mathrm{POS}$ values to only some 24 pixels which lacks information on the other pixels having polarization data, our estimation of the magnetic relaxation strength and the study of its effect on the RAT Alignment is also limited to these pixels only. There are other pixels showing high $P$ values of around 15\% but lacks $B_\mathrm{POS}$ information. However in these 24 pixels, there are some pixels with high $P$ values ranging from 5\% to around 8\%, mostly in the envelopes of the cores C2, C3 and C4 in the F13 filament and $6-9.5$\% near C6 and C7 cores in the F13S filament. The pixels near the cores C2, C3 and C4 are also associated with typical gas volume densities of around $6 \times 10^{3}$ $\mathrm{cm^{-3}}$. These observed $P$ values reaching around 8\% in the envelopes of the C2, C3 and C4 cores can be considered as high values and only RATs may not be able to produce these values. We also find some pixels having enhanced magnetic relaxation strength but showing lower $P$ values in the F13C region, associated with higher $a_\mathrm{align}$ values.

We explore the importance of enhanced magnetic relaxation strength on the RAT alignment of grains and investigate whether the observed high $P$ values in both the F13 and F13S filaments are purely due to RAT Alignment mechanism or M-RAT mechanism. When the magnetic relaxation occurs much faster than the gas collision damping, the magnetic relaxation strength is considered to be effective for the alignment of the grains. In the presence of both the suprathermal rotation of the grains by RATs and the enhanced magnetic relaxation strength, the RAT alignment efficiency of grains can be significantly increased (\citealt{2016ApJ...831..159H}; \citealt{2022AJ....164..248H}). Super-paramagnetic grains can achieve perfect alignment by the M-RAT mechanism when $\delta_\mathrm{mag,sp} > 10$ \citep{2016ApJ...831..159H}. \cite{2024arXiv240710079C} found that observed high $P$ values in the envelopes of protostellar cores could be produced by the thermal emission of the aligned super-paramagnetic grains with $N_\mathrm{cl} \approx 10^2-10^3$ which can produce high intrinsic polarization fraction. In our estimation of the magnetic relaxation strength in those 24 pixels in both the F13 and F13S filaments as shown in Figure \ref{Figure:mr_map}, a small level of iron inclusions can produce $\delta_\mathrm{mag,sp} >> 10$, especially in the F13N and F13C regions. However, we find $\delta_\mathrm{mag,sp} << 10$ in the F13S region. 

In the F13N region, we find that some pixels which are associated with higher $a_\mathrm{align}$ values as well as higher $\delta_\mathrm{mag,sp}$ values have higher $P$ values of around 6\%. Here, only RATs may not be able to produce these high $P$ values in these denser regions and the importance of alignment of grains by M-RAT mechanism can be significant. Again we see some pixels having lower $a_\mathrm{align}$ values but $\delta_\mathrm{mag,sp} > 10$ have lower $P$ values of around 4.6\% but still high. However, we expect these pixels to have higher $P$ values than the observed values. These smaller $P$ values may be due to the local dominance of magnetic field tangling along the line of sight over the grain alignment in these particular pixels.

In F13C region, we find that $P$ increases with the simultaneous increase in $\delta_\mathrm{mag,sp}$ and the decrease in $a_\mathrm{align}$ (see Figure \ref{Figure:mr_map}). This may imply the significant role of enhanced magnetic relaxation on the RAT alignment efficiency of grains. Some pixels which are associated with higher $a_\mathrm{align}$ values and found closer to the denser C2 core region, show smaller $P$ values even though the $\delta_\mathrm{mag,sp} > 10$. This may be due to the saturation of M-RAT alignment efficiency and the increase in the minimum alignment size of grains. 

% Also, the smaller $P$ values, even though the $\delta_\mathrm{mag,sp} > 10$, in some pixels associated with higher $a_\mathrm{align}$ values and found closer to the denser C2 core region may be due to the saturation of M-RAT alignment efficiency and the increase in the minimum alignment size of grains. 

% However, in the F13N region, we find that some pixels associated with higher $a_\mathrm{align}$ values as well as higher $\delta_\mathrm{mag}$ values have higher $P$ values of around 6\%. For this condition, only RATs may not be able to produce these high $P$ values in these denser regions and the importance of alignment of grains by M-RAT mechanism can be significant. Again we see some pixels having lower $a_\mathrm{align}$ values but $\delta_\mathrm{mag} > 10$ have lower $P$ values of around 4.6\% but still high. However, we expect these pixels to have higher $P$ values than the observed values. These smaller $P$ values may be due to the local dominance of magnetic field tangling along the line of sight over the grain alignment in these particular pixels. 

The enhanced magnetic relaxation combined with RATs can increase the alignment efficiency of grains and result in higher polarization fraction. Given the limitation of the data and the study to only some pixels and also use of estimated gas volume density by assuming cylindrical geometry of the filaments, a robust conclusion on the M-RAT mechanism can not be made. However, our study in these limited pixels provides us potential hints for the possibility of significant role of M-RAT alignment mechanism to explain the observed higher $P$ values in some regions of the F13N and F13C regions. 

For the F13S region which is associated with weak magnetic fields, since the $\delta_\mathrm{mag,sp} << 10$, the M-RAT mechanism is found to be not efficient in this region and the observed higher $P$ values are due to RATs only. The F13S filament is comparatively less denser than the F13 filament and is closer to the nearby B-type star BD+46. This region has higher dust temperature values compared to F13 region and hence lower $a_\mathrm{align}$ values. Therefore, in this F13S region the RATs may be strong enough to produce efficient alignment of grains in the outer less dense regions due to its close proximity to the star BD+46, resulting in high $P$ values. 

Our study further supports for the possibility of the M-RAT alignment mechanism \citep{2016ApJ...831..159H} in the F13N and F13C regions of the F13 filament. All the discussions made in Section \ref{section:Discussions_P-I-NH2}, \ref{section:Discussions_P-Td} and \ref{section:Discussions_P-a_align} altogether provide us strong evidence for the possible explanation of the observed depolarization in the denser regions of the filaments by the decrease in the RAT alignment efficiency of grains in the denser regions of these filaments. Also, from the discussion made in Section \ref{section:Discussion_magnetic_relaxation} we get potential hints for the M-RAT alignment mechanism of grains in some of the regions of the F13 filament.

\section{Conclusions}\label{section:Conclusions}
In this work, we investigate the dust grain alignment mechanisms in the F13 and F13S filamentary regions of the Cocoon Nebula (IC 5146) using thermal dust polarization observations from POL-2 instrument mounted on the James Clerk Maxwell Telescope (JCMT) at 850 $\mu$m. We study F13 region in two sub-regions as F13N and F13C that denote North and Center regions of this filament. Our main results are summarized as follows:
\vspace{0.3cm}

\textbf{1.} We find that the polarization fraction $P$ decreases with the increase in the total intensity $I$ and the gas column density $N(\mathrm{H_2})$, the so called polarization hole, in each of the F13N, F13C and F13S regions. The polarization fraction $P$ is found to increase with the increase in the dust temperatures $T_\mathrm{d}$ or equivalently the radiation field strength in each of the regions.
\vspace{0.3cm}

\textbf{2.} We investigate for any effective role of magnetic field tangling along the line-of-sight in causing the observed polarization hole by estimating the polarization angle dispersion function $S$ and the averaged grain alignment efficiency $P \times S$ and then analysing relations between $P$ and $S$; $P\times S$ and $I$; $P \times S$ and $N(\mathrm{H_2})$ and $P \times S$ and $T_\mathrm{d}$. We find that the effect of magnetic field tangling on the observed polarization hole is less and not much significant and this polarization hole is due to the decrease in RAT alignment efficiency of grains in the denser regions.
\vspace{0.3cm}

\textbf{3.} Further, to test whether the RAT-A mechanism can reproduce the observational results, we estimate the minimum alignment size of grains $a_\mathrm{align}$ for all the regions. We find that $a_\mathrm{align}$ increases with the increase in $I$ in each region, which is in agreement with the RAT-A theory. Again, from the analyses of the variations of $P$ and $P \times S$ with $a_\mathrm{align}$, we find that the magnetic field tangling has no significant effect on the depolarization in each of the regions. The decrease in $P$ and $P \times S$ with the increase in $a_\mathrm{align}$ can be well explained by the RAT-A theory.
\vspace{0.3cm}

\textbf{4.} From all the analyses, we find strong evidence for the RAT-A mechanism that can explain that the observed depolarizations in the denser regions of the F13N, F13C and F13S regions are due to the decrease in the RAT alignment efficiency of grains in the denser regions.   
\vspace{0.3cm}

\textbf{5.} We also study the importance of magnetic relaxation on the RAT alignment of grains by estimating the magnetic relaxation strength. Our study finds that the observed high polarization fractions in some of the regions of the F13 filament may be potentially due to the combined effects of both the suprathermal rotation of grains by RATs and the enhanced magnetic relaxation strength, supporting the possibility of the Magnetically-enhanced RAT (M-RAT) alignment mechanism of grains. The M-RAT mechanism is found to be not favorable in the F13S filament which has weak magnetic fields and the observed high $P$ values in the outer regions of this filament can be due to efficient alignment of grains in the outer less dense regions by the strong RAdiative Torques (RATs) from the nearby close B-type star BD+46.

%\begin{acknowledgments}
\section{Acknowledgements}

This research has made use of observation data from James Clerk Maxwell Telescope (JCMT) POL-2 instrument. JCMT is operated by the East Asian Observatory on behalf of the National Astronomical Observatory of Japan; Academia Sinica Institute of Astronomy and Astrophysics; the Korea Astronomy and Space Science Institute; the Operation, Maintenance and Upgrading Fund for Astronomical Telescopes and Facility Instruments, budgeted from the Ministry of Finance of China and administrated by the Chinese Academy of Sciences and, the National Key R and D Program of China. T.H. acknowledges the support from the main research project (No. 2025186902) from Korea Astronomy and Space Science Institute (KASI). P.N.D and N.B.N were funded by Vingroup Innovation Foundation (VINIF) under project code VINIF.2023.DA.057.

%\end{acknowledgments}

\vspace{5mm}
\facilities{James Clerk Maxwell Telescope, Herschel Space Observatory}

\software{Astropy \citep{2013A&A...558A..33A, 2018AJ....156..123A}; Scipy \citep{2020NatMe..17..261V}}

\bibliography{Cocoon_nebula_references}{}
\bibliographystyle{aasjournal}

\end{document}